\documentclass{article}

\usepackage{arxiv}

\usepackage[utf8]{inputenc} 
\usepackage[T1]{fontenc}    
\usepackage{hyperref}       
\usepackage{url}            
\usepackage{booktabs}       
\usepackage{amsfont s}      
\usepackage{nicefrac}       
\usepackage{microtype}      
\usepackage{lipsum}			
\usepackage{graphicx}

\usepackage{natbib}
\setcitestyle{numbers}
\setcitestyle{open={[},close={]}}

\usepackage{doi}
\usepackage{csquotes}
\usepackage{amsmath, amssymb}
\usepackage{setspace}
\usepackage[table,xcdraw]{xcolor}
\definecolor{table_background}{rgb}{0.933,0.933,0.933}

\usepackage{lscape}
\usepackage{pdflscape}

\newcommand{\paragraphNL}[1]{\paragraph{#1}\mbox{}\\}

\title{Group cohesion in multi-agent scenarios as an emergent behavior}

\author{ 
    \includegraphics[scale=0.06]{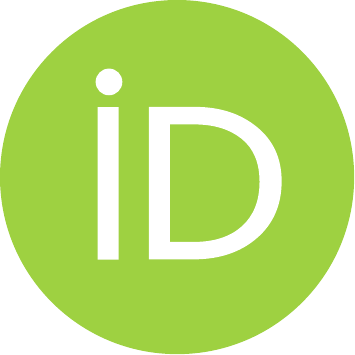}\hspace{1mm}Gianluca G. A. Volkmer \\
	Institute of Computer Science \\
	Freie Universität Berlin \\
	Arnimallee 7, 14195 Berlin \\
	\texttt{gianluca.volkmer@fu-berlin.de} \\
	\And
	\includegraphics[scale=0.06]{orcid.pdf}\hspace{1mm}Nabil Alsabah \\
	Institute of Computer Science \\
	Freie Universität Berlin \\
	Arnimallee 7, 14195 Berlin \\
	\texttt{nabil.alsabah@fu-berlin.de} \\
	}

\renewcommand{\shorttitle}{Emergent Group Cohesion}

\hypersetup{
pdftitle={Group cohesion in multi-agent scenarios as an emergent phenomenon},
pdfsubject={multi-agent systems},
pdfauthor={Gianluca G. A. Volkmer, Nabil Alsabah},
pdfkeywords={Multi-Agent Systems, Cognitive Architectures, PSI Framework},
}

\date{}

\begin{document}

\maketitle

\begin{abstract}
  In this paper, we elaborate on the design and discuss the results of a multi-agent simulation that we have developed using the PSI cognitive architecture. We demonstrate that imbuing agents with intrinsic needs for group affiliation, certainty and competence will lead to the emergence of social behavior among agents. This behavior expresses itself in altruism toward in-group agents and adversarial tendencies toward out-group agents. Our simulation also shows how parameterization can have dramatic effects on agent behavior. Introducing an out-group bias, for example, not only made agents behave aggressively toward members of the other group, but it also increased in-group cohesion. Similarly, environmental and situational factors facilitated the emergence of outliers: agents from adversarial groups becoming close friends. 
  
  Overall, this simulation showcases the power of psychological frameworks, in general, and the PSI paradigm, in particular, to bring about human-like behavioral patterns in an emergent fashion.
\end{abstract}


\section{Introduction}
The integration of artificial intelligence into multi-agent systems (MAS) has garnered an ever increasing attention among AI researchers. Of special interest are agents that exhibit social behavioral patterns, such as coordination, cooperation and conflict resolution \cite{Alshabi2007}. Thereby, researchers have relied on classical approaches \cite{OnCooperationInMAS, ConflictResolutionMAS, DesigningAndUnderstandingAdaptiveGroupBehavior}. In recent publications, researchers used reinforcement learning to train their agents in multi-agent environments \cite{EmergentToolUse, HumanLevelMPGamesRL, StarCraftMASRL}. This subfield of machine learning allows agents to deduce optimal behavior solely from the problem formulation using a reward function that gives (delayed) feedback on actions. Social behavior, such as cooperation, eventually emerges as agents optimize their behavior to reach a predefined goal.

Using AI frameworks that explicitly integrate psychological insights into human behavior might seem as a viable alternative when designing social multi-agent systems. In fact, over the course of the last four decades, cognitive scientists have developed so-called cognitive architectures to provide unified models of cognition and serve as frameworks for introducing human-like behavioral patterns into AI agents \cite{Kotseruba2016, Laird1991}. Some cognitive architectures, like Soar \cite{Laird2019, Newell1994} and Icarus \cite{Konik2009} are purely symbolic: They emulate aspects of planing and reasoning through the vehicle of production rules. Others feature subsymbolic architectures that codify declarative and procedural knowledge in connectionist networks \cite{Banich2018, Churchland1988}. Still others are hybrid systems that aspire to match human cognitive abilities through a combination of symbolic and subsymbolic paradigms, with ACT-R being the most prominent representative of this school \cite{Anderson2004, Zylberberg2011}.

Yet despite their proclaimed aim of providing a framework for replicating aspects of human intelligence and behavior, cognitive architectures have not been very successful in this regard. This maybe due to three constraining factors. (1) The starting point of cognitive architectures is--as the name implies--human cognition, not human motives. Cognitive frameworks model cognitive skills, like perception and object recognition. They venture into more advanced mental abilities, like planning and reasoning. They aspire to unravel the mysteries of semantic understanding. However, cognitive agents have no built-in mechanisms for developing wants, desires, intentions, and preferences that are tied to a need structure. (2) Mainstream cognitive architectures view the mind as a collection of specialized modules each dedicated to solving a particular set of tasks \cite{Carruthers2008}. Even emotions are regarded as a module which framework architects would map out at some point in the future \cite{Ritter2019}. (3) Cognitive frameworks ignore the fact that complex human behavior does not arise in a void. Rather, it emerges through the interaction of motivational, cognitive, and emotional factors—at the personal and the group level—with a dynamic and complex environment \cite{McAdams2008, Reynolds2010}.

The PSI framework--outlined here in \ref{sec:agents}--has been designed specifically to address those shortcomings. We applied the PSI framework to design a multi-agent simulation. Our aim was to address three sets of questions:

\begin{enumerate}
        \item How does intra- and inter-group social dynamics play out when you have a society of need-driven agents?
        \item How does segmenting agents along group lines affect social cohesion? What other changes to the social structure occur due to group segregation?
        \item Do inter-group relationships develop? If so, under which circumstances?
    \end{enumerate}
    
In this paper, we will present our findings. But first we will discuss the environment as well as the agent design in \ref{sec:TheSimulation}. Then, in \ref{sec:ResultsAndAnalysis} we will put forward and elaborate our results. Finally, in \ref{sec:ConclusionAndFutureWork} we will close the paper with a short discussion.

\section{The Simulation}\label{sec:TheSimulation}
\subsection{The Environment}
We have used the game engine \textit{Unity} to develop our multi-agent simulation.\footnote{You can download the simulation from \cite{GianlucaGithub}.} The simulation consists of a discrete hexagonal map in which agents take actions at each time step $t$. A hexagonal map provides an optimal plane tiling since hexagons preserve equal distance in all directions.
        
The map of the simulation consists of an accessible region called \enquote{grassland} and three inaccessible regions called \enquote{mountain}, \enquote{forest}, and \enquote{ocean}. Inaccessible terrain is there only for aesthetic reasons. It does not influence the simulation in any way.

\begin{figure}[t]
	\centering
	\includegraphics[width=0.8\textwidth]{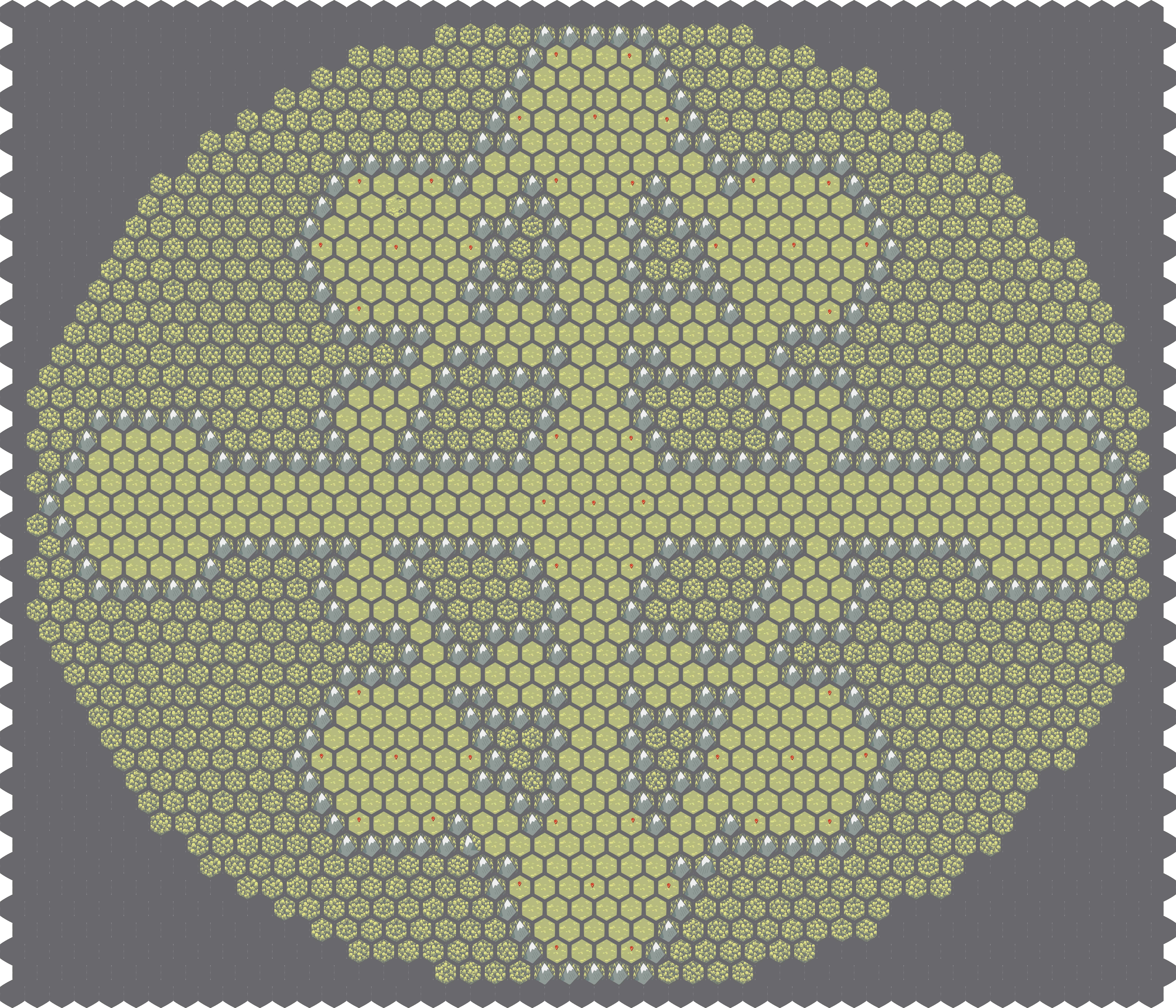}
	\caption{The environment that was used in the simulation. On the left, we have the spawn area of agent group one, and on the right, the spawn area of agent group two. We can see the contested area in the middle. The food spawns are marked with a small red point. Agents are free to traverse the entire map.}
	\label{fig:SimulationEnvironment}
\end{figure}

The accessible region is divided into three subregions: The spawn area of agent group one, a contested area, and the spawn area of agent group two. The spawn areas serve as the home territories for agent groups. They only contain agent spawn points. It is the contested area where agents come together to search for food and interact with each other. It is important to mention that agents have access to the entire map as boundaries are not strictly enforced. Figure \ref{fig:SimulationEnvironment} provides a snapshot of the simulation.
        
As described in the next chapter, agents are intrinsically motivated to forage. The map has food spawn points. Food randomly appears there at a rate determined by a given probability $\alpha_f$. After some trial and error, we settled on an $\alpha_f = 0.02$. 

\subsection{The Agents}\label{sec:agents}
\subsubsection{Need Structure}\label{sec:NeedStructure}
Our agents are modelled according to the \textit{PSI Theory}. This theory is a framework for designing AI agents that can achieve human-level behavioral complexity. It has been developed by Dietrich Dörner and his group at the University of Bamberg for over 40 years. PSI stands for person-system interaction. The framework has been discussed in several key publications \cite{MicroPsi2, EinfuhrungAffiliation, DornerComputationalArchitecturOfCognition, BauplanFurEineSeele}.

The PSI theory postulates that living organisms are self-organizing systems that engage in goal-directed behavior in order to balance a number of internal, critical variables. Those critical variables represent the system’s needs. Humans, for example, are programmed by evolution to, among others, avoid pain, breath, drink, and eat, engage in sexual activities, seek attachment and certainty, and learn to deal with arising challenges in their environment.\footnote{ Psychological literature identifies several paradigms for clustering human needs. The psychoanalytic school reduces human motives to sexuality and aggression \cite{Freud1955}. The humanistic school argues that humans strive to become self-determining organisms \cite{Maslow1968, Rogers1959}. And the diversity school sees humans as being motivated by many different factors that vary from person to person \cite{Murray2007}.}

The agents in our simulation come with five basic needs:
\begin{enumerate}
    \item \textit{Pain-Avoidance}: The need for physical and mental well-being.
    \item \textit{Energy}: The need for food.
    \item \textit{Affiliation}: The need for togetherness, belonging, and being part of a group.
    \item \textit{Certainty}: The need for safety, security, and predictability.
    \item \textit{Competence}: The need for feeling capable of dealing with situational challenges. Its subcategories include control, power, self-confidence.
\end{enumerate}

\begin{figure} [h]
    \centering
    \includegraphics[width=\textwidth]{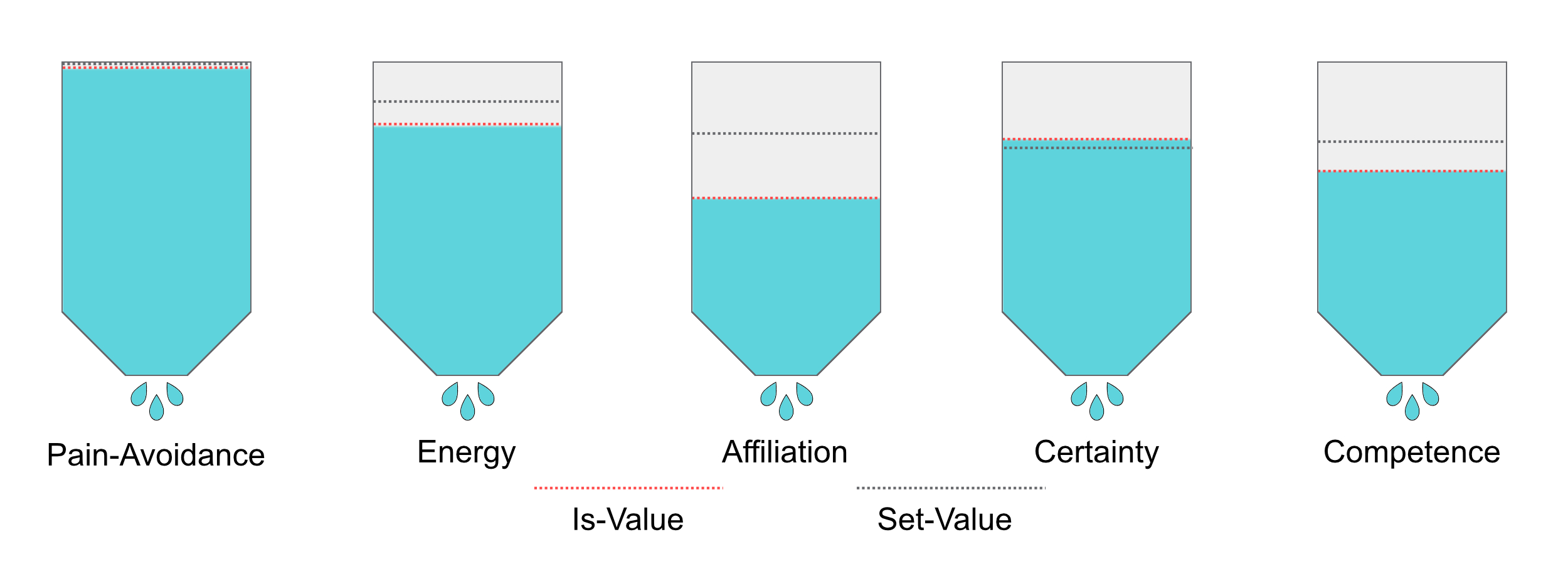}
    \caption{We can think of needs as storage tanks. Each time step $t$ the current value is updated according to the leakage value. Agents are intrinsically driven to engage in behavior directed toward reducing deviations between current and set values.}
    \label{fig:NeedTanks}
\end{figure}

Metaphorically speaking, the PSI architecture conceptualizes needs as storage tanks, each with a \textit{current value} and a \textit{set value} (see figure \ref{fig:NeedTanks}). The storage tanks are leaky. At each time step $t$, the current value $current^{need}$ is adjusted according to the leakage value $\eta^{need}$ of the respective need tank:

\begin{equation*}
    current^{need}_{t} = current^{need}_{t-1} - \eta^{need}
\end{equation*}

Thus, leakages lead to deviations to occur between the current value and the set value. Whenever that happens, the PSI agent engages in goal-directed behavior in order to restore balance to its internal system by eliminating or, at least, reducing the deviation. Further, the process of need satisfaction—e.g. food consumption—produces \textit{pleasure signals} while going through an aversive experience yields \textit{displeasure signals}. Those signals increase and decrease the current value in storage tanks respectively.

The size of the leak—or, the decay value—determines how fast a storage tank empties. The leak size can vary from storage tank to storage tank, and from agent to agent. This gives agents the first shade of personality: Those with a relatively large leak in the affiliation tank will engage more often in pronounced social behavior; those with a large leak in the energy intake tank will be addicted to energy intake; and those with a large leak in the competence tank will be the adventurous type, seeking challenges for the sake of overcoming them.\footnote{ In this simulation, we determined the leakage values according to a Gaussian distribution. The specific $\mu$ and $\sigma$ values were chosen through trial and error. For more information, please refer to appendix \ref{sec:SimulationSettings}. The specific set values and leakage values of the agents used in our simulation can be found in the appendix \ref{sec:AgentPersonalityValues}.}

\subsubsection{Memory}
An integral part of intelligent behavior is the ability to learn and remember information about the environment and its inhabitants. To model this, we have created two separate memory systems: One pertains to remembering locations and one concerns modeling social relationships.
            
\paragraphNL{Location Memory}
At the start of a simulation cycle, the agents know nothing about the environment and therefore have to gather information about their surroundings. When they discover new hex tiles, they update their mental map of the world, here modeled as a simple graph of so-called \enquote{Memory-World-Cells}.
                
Agents could potentially experience pleasure and displeasure signals at each tile. Thus, each Memory-World-Cell also stores this information. So each time an agent experiences a new signal—e.g. an energy signal from consuming food—the agent updates the need satisfaction value $\nu^{need}$ associated with this specific Memory-World-Cell. Further, the positive (or, negative) association the agent has with a Memory-World-Cell extends to neighboring Memory-World-Cells as well—albeit with decreasing signal strength. This means that the farther away a tile is located from the signal emitting tile, the less it is associated with that particular signal. 
For this simulation, we have decided that the maximum number of affected tiles should be 4. So the need satisfaction association $\nu^{need}$ is updated as follows for a signal $\sigma^{need}$ and a distance $d$:
                
\begin{equation*}
    \nu^{need}_{i+1} = \nu^{need}_{i} + \sigma^{need} \cdot \max \Big( 0, \frac{4 - d}{4} \Big)
\end{equation*}
                
The need satisfaction association values are then clipped so that $\nu^{need} \in [-1; 1]$.
                
Forgetting is an integral mechanism of the memory system. Thus, our agents not only learn new information, but also forget old ones too. Each time step $t$ the need satisfaction association $\nu^{need}$ is updated by a forgetting rate $\varphi_{p}^{l}$ for positive associations and a forgetting rate $\varphi_{n}^{l}$ for negative associations. So, the need satisfaction association $\nu^{need}$ is updated as follows:
                
\begin{equation*}
    \nu^{need}_{t} =    
        \begin{cases}
            \nu^{need}_{t-1} \cdot \varphi_{p}^{l} \text{, if } \nu^{need}_{t-1} \geq 0 \\
            \nu^{need}_{t-1} \cdot \varphi_{n}^{l} \text{, if } \nu^{need}_{t-1} < 0
        \end{cases}
\end{equation*}
                
Cognitive psychology studies have demonstrated that humans forget negative events faster than positive ones \cite{TheFaidingBais}. Therefore we have chosen $\varphi_{n}^{l} < \varphi_{p}^{l}$. The exact values can be found in the appendix \ref{sec:SimulationSettings}.
                
\paragraphNL{Social Memory and Social Scores}
We have integrated a social score system into our simulation to model the mechanisms of social relationships. Agents keeps track of other agents they have met. Each connection has a social score $s \in [-1; 1]$, describing an agent's opinion of another agent. Agents with social scores $> 0$ are regarded as friendly, while scores $< 0$ are regarded as hostile.
        
When an agent meets another agent for the first time, they assign it a social score according to a Gaussian distribution. This function takes the new agent's group affiliation into account. Agents in the same group get a positive initial social score, while agents in the opposite group get a negative initial social score. The specific values for $\mu$ and $\sigma$ can be found in the appendix \ref{sec:SimulationSettings}.
Each time two agents interact socially--for example, when they exchange information or attack each other--they receive a positive or negative social score feedback.
                
Besides serving as an indicator of the strength of a relationship between two agents, social scores also influence the decision process of the agents. The reward and punishment for social and anti-social behavior is adjusted according to the social score between the two interacting agents. For example, if an agent attacks a friend, the current value in the affiliation tank takes a large hit. Attacking an adversarial agent, on the other hand, results in little to no affiliation loss.
                
Similar to the forgetting rate for location information, there is also a forgetting rate for social information. This allows us to model the decay of relationships over time. The forgetting rate for positive relationships is defined as $\varphi_{p}^{s}$ and for negative relationships as $\varphi_{n}^{s}$. The mechanism works analogously to the forgetting mechanism for location information. 
                
\subsubsection{Action Plans} \label{sec:ActionPlans}
Agents have a set of predefined action plans each of which specifies a permissible behavior. In this simulation, agent behavior has been hard-coded. It is though possible to expand the simulation so to allow agents to rely on classical AI planning algorithms or reinforcement learning methods instead. 
            
This section offers an overview of our action plans. First, we will give a general introduction to the structure of an action plan. Then, we will define all basic actions that serve as the fundamental building blocks of agent behavior. Finally, we will describe specific action plans that agents can choose from. The selection process is then described in detail in section \ref{sec:MotiveSelection}.

When an agent learns a new action plan $a$, it initializes it with a success probability $Pr_{a} = 1$ and an expected need satisfaction value $\mathbb{E}_{a}^{need}(\theta)$. The variable $\theta$ reflects the internal state of the agent and it contains a description of the concrete situation the agent is in. Successfully executing an action plan generates a positive need satisfaction signal. Failing to reach the goal specified by the action plan creates a negative need satisfaction signal. The agent uses this signal to update the expected reward and punishment $\mathbb{E}_{a}^{need}(\theta)$ for that specific action plan. Simultaneously, the agent also updates the expected success probability $Pr_{a}$. The success probability plays, as we shall see, an essential role in the motive selection process.
In addition to the expected need satisfaction value and the success probability, each action plan $a$ also has an urgency value provided by the function $urgency_{a}(\theta)$. This value varies depending on the current state of the agent $\theta$.

The urgency value affects--but does not determine--which action plan is currently being pursued by the agent. For example, image that agent A is currently executing an action plan to explore a hitherto unexplored part of the map. While exploring, it accidentally finds food. Even though the agent's need for certainty is higher than its need for energy intake, the specific circumstances of the situation--i.e., there is food in the immediate vicinity--incentivizes the agent to interrupt exploration and gather the food it has just discovered. This incentivization happens by increasing the urgency of the action plan \textit{collect\_food}.

\paragraphNL{Basic Actions}
Each action plan is made up of a set of basic actions. These actions include walking, calling out other agents in the region, collecting and eating food, and attacking nearby agents.
                
When agents want to go to a specific hex tile, they calculate a path based on the tiles they already know. If another agent blocks their way, they have a 50\% chance of starting a collision avoidance maneuver. This guarantees that no deadlocks happen.
                
The shout-out action is used to inform nearby agents of the agent's internal state. There are five types of \enquote{shout-outs}: A cry for help when the agent was attacked, a general request for food, a general request for healing injuries, a request to exchange social information, and a request to exchange location information. The call for help and the calls for knowledge exchange contain additional information about who the request is regarding. Each call is broadcasted to all agents in viewing range of the agent, i.e., four tiles. Incoming requests are processed within the agent's \enquote{tick method} (see \ref{sec:TickMethod}).
                
The collect and eat actions simply handle the gathering and consumption of food from the current hex tile or the food storage of the agent. The hit action deals with inflicting damage on other agents. The amount of damage the other agent receives from the hit is determined by a Gaussian distribution. The parameters $\mu$ and $\sigma$ of the distribution can be found in the appendix \ref{sec:SimulationSettings}.
                        
\paragraphNL{The Explore Action Plan}
Agents have the option of exploring unknown parts of the map. At the beginning of the simulation, the agent can enact this action plan to roam over hex tiles. If no unexplored tiles are left, the agent would pick a random tile to visit weighted by the associated certainty value. Tiles with lower associated certainty values are then preferred. This action plan expands the agent's knowledge of its environment, and thereby decreases the level of environmental uncertainty.
                
\paragraphNL{Food Related Action Plans}
The agents have three ways to get food: Go on a search for food and collect it from a nearby hex tile, look for and collect food from a known food cluster, and request food from nearby agents.
                
If the agent has not yet discovered or has failed to find food at a known food cluster, it can forage for food sources. Here it would select a random location from the list of explored locations stored in its memory. If the agent happens upon a food source in the selected tile or on the way to the selected tile, it would consume it. Otherwise, a new hex tile would be selected, and the process starts anew.
                
Every time agents happen upon an energy source, they update their mental map of food clusters. For each food cluster, they create an action plan with the middle hex tile as the center of the cluster. Now, if an agent gets hungry or just wants to collect food for later consumption, it could choose one of the known food clusters and go there. If it reaches the middle hex tile without finding any food, the action plan would fail, and the agent would have no other action plans to satisfy its need for energy.
                
As a last resort, agents would have to ask fellow agents for help and hope that nearby agents would have food in their storage and be willing to share it. For this, they send out a food request to all agents and then wait for five time steps. The action plan fails if other agents decide to ignore the plea for help. What incentive do agents have to comply with help requests? Well, helping others does produce affiliation signals that increase the current value in the affiliation tank. Thus, agents have an intrinsic need to act altruistically. 
            
\paragraphNL{Social and Anti-Social Action Plans}
A corner stone of this simulation is interactions among agents. In interacting with each other, agents have a repertoire of social and anti-social action plans they can choose from. Social behavior includes acknowledging each other, exchanging location information, gifting food, and healing injured agents. Social behavior generates affiliation signals. It also affects inter-agent relationships. Agents can also engage in anti-social behavior which generates anti-affiliation signals and also has negative effects on agent relationships. The most prominent act of anti-social behavior is attacking other agents. It is important to note that expected and actual need satisfaction in social situations is affected by the relationship of those agents interacting with each other.

Consider this scenario: Agent A and agent B meet up to exchange information about agent C. Thereby, one of the two agents would exchange the social score it has assigned to agent C $s^{\prime C}$ with the other agent. If the receiving agent does not know agent C already, it would add it to its memory. However, it would not just accept the sending agent’s opinion of agent C, as it were. Rather, it would apply a discount factor $\alpha_{s_1}$:

\begin{equation*}
    s^{C} = \alpha_{s_1} s^{\prime C}
\end{equation*}
                
If agent C is known to the receiving agent, then the receiving agent would update the social score it has assigned to agent C with a discount factor $\alpha_{s_2}$:
                
\begin{equation*}
    s^{C} = (1 - \alpha_{s_2}) s^{C} + \alpha_{s_2} s^{\prime C}
\end{equation*}
                
If two agents A and B decide to exchange location information, they would meet up and exchange need satisfaction associations for 10 random hex tiles each. Suppose the receiving agent knows one of the hex tiles it has just received from the other agent. In that case, it would update its need satisfaction association for that tile also by applying a discount factor of $\alpha_{l}$.
                
If an agent A calls out for food and agent B, with food in its storage, picks up the distress call, then both agents could initiate a food transfer from agent B to agent A. For this, agent B would navigate to agent A's location and hand over some food.
                
If agent A is hurt and agent B would like to help out, then both agents would activate the healing action plan. Agent B would go to agent A and heal $h$ health points. $h$ is determined by a Gaussian distribution. Agent A could also call out to be healed, resulting in an increased urgency, which might increase agent B's motivation to help.
                
Each of the above-mentioned action plans positively influences the social score. For this simulation, we decided to use $0.1$ as a social reward. Further, agents receive a negative social reward when they attack each other. The decision to engage in aggressive behavior is influenced by an agent's level of certainty and competence as well as the anti-affiliation signals that would be generated as a result. Hurting a friend comes with a higher anti-affiliation price tag than injuring strangers. Aggressive behavior not only comes with affiliation risks, but it also endangers the physical well-being of the aggressor. Therefore, agents are thereby structurally discouraged from taking aggressive actions--especially, when their health level is low.

If an agent decides to attack, it would seek out its target and strike. The attacked agent would register the attack, decide whether it would prefer to fight or flee, and adjust the social score associated with the attacker by a value of $-0.1$. Should the victim decide to flee, it would pick out a random \textit{safe} location from its location memory. If the agent it is fleeing from is out of range, the action would then be counted as successful. If the fleeing agent should call out for help and another agent would respond positively, then the helping agent would get a positive social reward.

\subsubsection{Motive Selection} \label{sec:MotiveSelection}
A PSI agent is a multi-stable system. This means that it seeks to achieve multiple stable equilibrium points regarding the agent’s needs. Satisfying one need might increase another. And since several need deviations might exist simultaneously and since there are several available or conceivable action plans to satisfy any given need, the agent must have an internal mechanism for deciding on (1) which need it wants to satisfy and (2) which action plan it wants to enact. 
            
Let us start with the first decision the agent has to make: The agent should figure out how important each need satisfaction is. Relying on deviations between current value and set point is not adequate. Energy intake, for instance, should have a higher priority than, say, socializing. Thus, we have to weigh the need strength by multiplying it with a constant. The concrete values used in this simulation can be found in table \ref{tab:NeedWeights}.

\begin{table}[h]
\centering
\begin{tabular}{|l|c|}
\hline
\multicolumn{1}{|c|}{\textbf{Need}} & \multicolumn{1}{l|}{\textbf{Weight $w$}} \\ \hline
	Pain-Avoidance             & 3		\\ \hline
	Energy                     & 2      \\ \hline
	Affiliation                & 1		\\ \hline
	Certainty                  & 1		\\ \hline
	Competence                 & 1		\\ \hline
\end{tabular}
\caption{Modeling the hierarchy of needs in the need-indicator $n^{need}$ using weights $w$.}
\label{tab:NeedWeights}
\end{table}

Putting everything together, we are now able to calculate the need indicator for each individual need:

\noindent
\begin{equation*}
	n^{need} = \max(0, set^{need} - is^{need}) \cdot w^{need}
\end{equation*}

Then, the agent will search its action plan memory for adequate action plans for pursuing each individual need. As in psychology, we also term the combination of a need $n$ and an action plan $a$ as a \textit{motive}. Each motive has an associated motive strength $m$. We can now use the previously calculated need indicators $n^{need}$ as well as the expected need satisfaction of the action plan $a$ $\mathbb{E}_{a}^{need}$ to calculate the preliminary motive strength:

\begin{equation*}
m^{\prime} = \sum_{i \in Needs} n^{i} \cdot \mathbb{E}_{a}^{i}(\theta)
\end{equation*}
    
The agent should also take into account the success probability $Pr_{a}$ of an action plan $a$. Further, the agent must consider its ability in tackling arising challenges. The current level in the competence tank $is^{competence}$ is an adequate proxy for the agent's general competence level. Thus, we introduce the competence indicator $c_{a}$ as a variable that brings those two components together:
            
\begin{equation*}
c_{a} = (1 - \alpha_{c}) \cdot is^{competence} + \alpha_{c} \cdot Pr_{a}
\end{equation*}
            
The discount factor $\alpha_{c} \in [0; 1]$ controls how much weight is given to the current value in the competence tank as well as the success probability $Pr_{a}$, respectively. For this simulation we have decided to set both variables equally at $\alpha_{c} = 0.5$.
            
The final motive strength $m$ is then calculated using the preliminary motive strength $m^{\prime}$, the current $urgency_{a}(\theta)$ of the action plan, and the competence indicator $c_{a}$:
            
\begin{equation*}
m = \Big( m^{\prime} + urgency_{a}(\theta) \Big) \cdot c_{a}
\end{equation*}
            
Finally, the agent will pick the motive--and with that, the action plan--with the highest motive strength.

Motives are dynamic. Need indicators change all the time due to tank leakage, upcoming urgencies, and enacting action plans with aversive effects. Competence indicators also change due to the agent’s successes and failures while acting in a dynamic and complex environment. The dynamic nature of motive strengths and the structure of the motive selection mechanism presents us with a problem: the agent might experience motive fluttering as it cycles from one motive to the next. To avoid the fluttering effect, we introduce a threshold $\tau$ to protect the current motive. To replace the current motive, the motive strength $m_{j}$ of a rival motive has to be larger than the motive strength $m_{i}$ of the current motive plus the threshold $\tau$.
For this simulation, we have found that $\tau = 0.02$ is sufficient to avoid behavior fluttering.
            
\subsubsection{The Tick Method} \label{sec:TickMethod}
Each time step, our Tick Method is called. This method is the brain of our PSI agents. In summary, this method does the following:
        
\begin{enumerate}
	\item Update the need tanks according to the leakage values.
    \item Forget information about location and social information.
    \item Sense the environment and all agents in the field of view.
    \item Process all incoming requests, e.g., calls from other agents for food or help.
	\item If needed, run the motive selection process to confirm if the current motive should still remain the action-guiding motive or if a rival motive should take over.
	\item If the agent has 0 energy: Deduct one point of health to simulate starvation.
\end{enumerate}

\section{Results and Analysis} \label{sec:ResultsAndAnalysis}
\subsection{Overview} \label{sec:TheScenarios}
This section will introduce the three scenarios we have implemented so to answer the defined research questions. We have created three different pairs of teams using the random seeds 1, 2, and 3. The specific set- and leakage-values that define the agent's personality can be found in the appendix \ref{sec:AgentPersonalityValues}.
All scenarios were then executed for all group pairs using the simulation seeds 1-3. In total, this equals to 9 different runs per scenario.
    
\paragraphNL{Scenario 1 - A Single Group}
In this scenario, we will look closely at the agents' behavior when no adversarial group is present. We use this scenario as a baseline for comparing the intra- and inter-group behavior with the other two scenarios. We predict that the agents will engage in intra-group fighting and, because of that, a social split inside the group will take place.
    
\paragraphNL{Scenario 2 - Two Highly Adversarial Groups}
In this scenario, we introduce the second group of agents. The social scores of members in opposing groups are initialized with very low values. The concrete initial social score values are picked using a Gaussian distribution with $\sigma = -0.9$ and $\mu = 0.05$.
        
We believe that introducing a second group of agents with low social scores will redirect the conflict toward the out-group. This should result in a stabilization of social scores for in-group members. Relationships between the two groups of agents should remain bad.
        
\paragraphNL{Scenario 3 - Two Moderately Adversarial Groups}
In this last scenario, we look at a variation of scenario two, where we adapt the social scores to have a moderate low value. The values for this scenario are also given a Gaussian distribution with $\sigma = -0.5$ and $\mu = 0.05$.
        
In this scenario, we hope to find a mixture of behavior from scenarios one and two. The tension between the two groups should still be visible with slight relaxation between some agents. In addition, we hope to find occasional links between members of the two opposing groups, similar to Romeo and Juliet from the eponymous drama.

\subsection{Scenario 1: One Group} 
In all runs of our simulation, we observed two main patterns of behavior that were prevalent at the beginning: An exploratory approach and an engage-heal approach. Both agent groups that started the simulation by exploring their surroundings were lacking in both certainty and competence. All other needs were satisfied.
        
The group of agents that chose to attack neighboring group members did not have any shortage in the certainty tank. Therefore, they only tried to maximize their gain of competence. As we have calibrated the simulation to offer the greatest reward of competence for a successful attack and as the other needs were already satisfied, the logical decision for an agent was to attack its neighbor. However, the agents quickly realized that this tactic backfires: It leads to a significant loss in affiliation and a loss in pain avoidance as other agents might decide to defend themselves. The action-guiding motive after this becomes healing injuries: As multiple agents were attacked, we had several agents with low health levels. Healing them leads to an increase in affiliation. This behavior is also visible in all other scenarios.

\begin{figure}[h]
	\centering
	\includegraphics[width=0.8\textwidth]{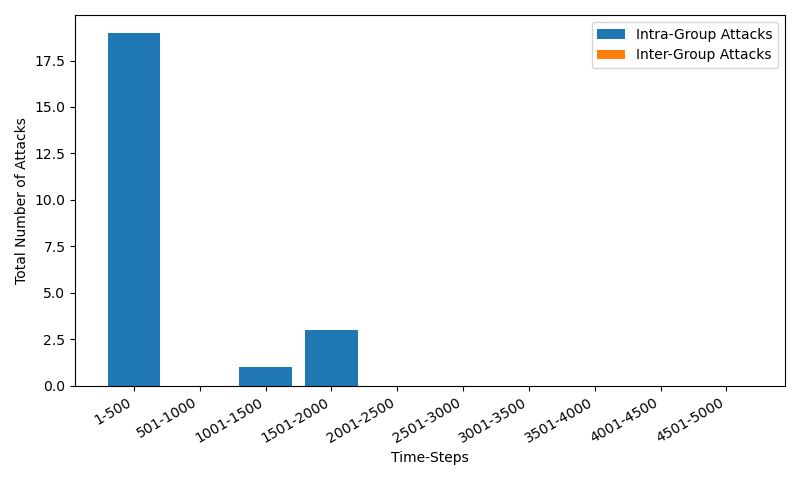}
	\caption{Counts of engage activities in bins of 500 time steps in scenario 1 with group 2 and seed 1. Note the majority of in-group attacks happens at the beginning of the simulation. After that, violent behavior becomes rare. Please refer to \ref{fig:s1_engage_actionplan_summary} for a summary of all graphs describing the engage action in scenario 1.}
	\label{fig:s1g2s1_engage_chart}
\end{figure}

\begin{figure}[]
	\centering
	\includegraphics[width=0.8\textwidth]{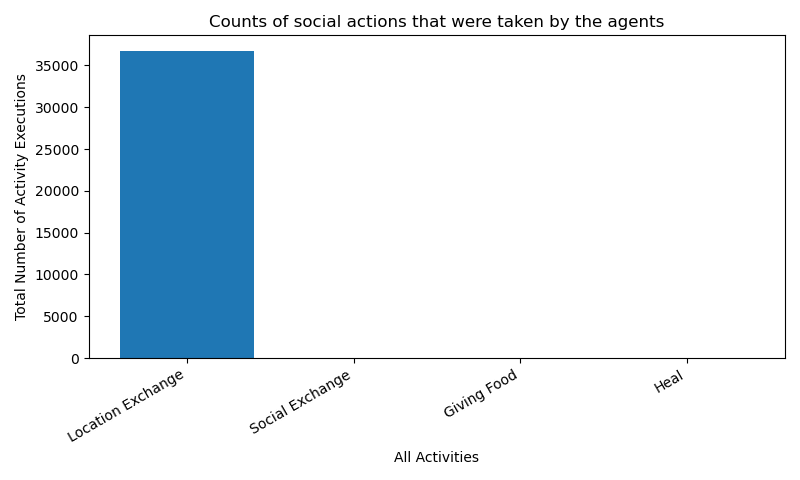}
	\caption{Counts of all action plans that were chosen throughout the whole simulation of scenario 1 with group 2 and seed 1. As we can see, agents chose to execute the exchange of location information action plan most of the time. Other social action plans are rarely executed. For a complete summary please refer to figure \ref{fig:s1_social_action_summary} in the appendix.}
	\label{fig:s1g2s1_social_action_count}
\end{figure}
        
In general, we could see that the choice to engage in violence is clustered at the beginning of the simulation (see figure \ref{fig:s1g2s1_engage_chart}). After that, the agents prefer social action plans. To be precise, we found out that the agents pick the exchange of location information action plan significantly more often than any other social action plan (see figure \ref{fig:s1g2s1_social_action_count}). However, it is unclear why they chose to do this instead of exchanging social information as they all would generate similar need satisfaction signals.
        
To understand the occasional outbursts of violent behavior in later parts of the simulation, we first have to understand the primary emerging strategy: After a few time steps in the simulation, the agent starts to get hungry and begins searching for food. Locations in which agents have found food are marked as food clusters and can therefore be revisited the next time they get hungry.

Now that the agent has had enough food, the main needs to satisfy become affiliation, certainty, and competence. To satisfy all three needs, the optimal course of action is to exchange information. The agent picks out an agent it has a close relationship with and start interacting with it. As this usually happens directly after satisfying the need for energy intake, the two agents will most likely be in or close to a food cluster. This motive is then executed again and again until one of the two agents gets hungry. The hungry agent then goes to the nearest food source to eat while its friend follows. As soon as the hunger is sated, the social interaction continues. In this scenario, the only situation where this setup breaks up is when one of the agents gets hungry, but no food can be found in the nearby cluster. The agents then decide to split and go their separate ways.

This behavior could lead to an unstable situation where tension could arise. Imagine the following scenario: Agent A and agent B follow the strategy described above and are currently in the information exchange phase. Another agent C, with a need for affiliation, joins them. C tries to talk with one of the two agents, but gets rejected. This rejection leads to a loss of certainty and competence. If agent A and agent B reject agent C often enough, the competence value drops below a tipping point. Agent C then could decide to take revenge for all the rejections its has suffered. This emerging behavior can be interpreted as a mixture of frustration and envy.
        
\begin{figure}[]
	\centering
	\includegraphics[width=\textwidth]{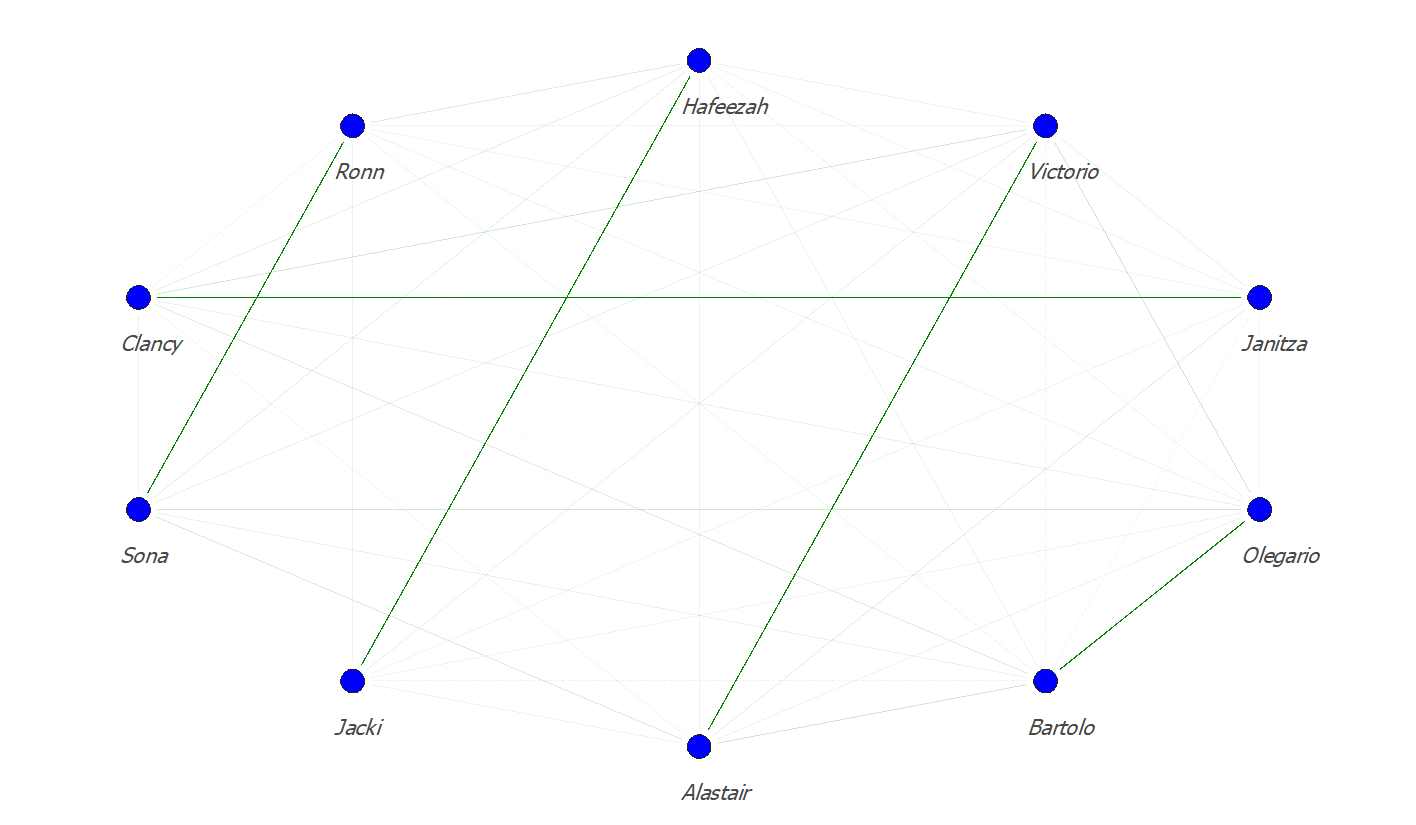}
	\caption{The final social network at $t=5000$ for scenario 1 with group 2 and seed 1. As we can see, the agents have split up into pairs of two with some weak connections to other agents. Some of the connections are also slightly negative. The split into pairs happens because of this behavioral pattern: Short food breaks between social interactions with their \enquote{best-friend}.}
	\label{fig:s1g2s1_social_graph}
\end{figure}
        
Finally, we wanted to see whether a social divide occurs \textit{within} the same group of agents. As you can see in figure \ref{fig:s1g2s1_social_graph}, the agents split up into groups of two. This is the result of an emerging behavior. A divide between agents of the same group is not directly visible in the figure. However, it occurs nonetheless throughout the violent behavior observed at the start of the simulation and the special cases described in the paragraph above. This leads to a temporary strong negative social score between the attacking and the attacked agent. As the relationship of the two agents decays over time, the social score of agents that have not interacted with each other for an extended period goes down to 0. This explains the strong disconnect of the group and the non-visible negative relationships.

\subsection{Scenario 2: Two Highly Adversarial Groups}
\begin{figure}[h]
	\centering
 	\includegraphics[width=0.8\textwidth]{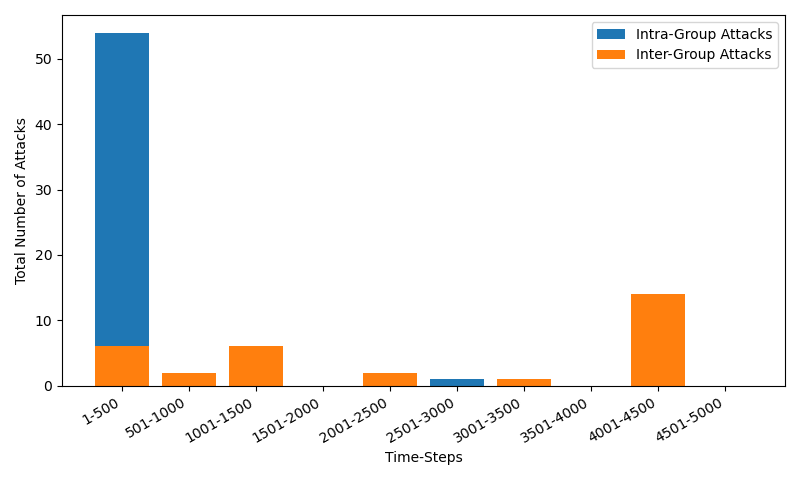}
	\caption{Stacked counts of all engage actions in scenario 2 with group 2 and seed 1. The blue bar shows the counts of intra-group fighting, while the orange bar shows the counts for inter-group fighting. We can see that most of the intra-group fighting happens at the beginning of the simulation. After the violent behavior is directed toward members of the opposing group. All counts can be found in the appendix in figure \ref{fig:s2_engage_actionplan_summary}.}
	\label{fig:s2g1s2_engage_summary}
\end{figure}

Over all runs of the second scenario, we observed an increase in violent behavior and a redirection of engage actions toward members of the adversarial group. An example run with group 2 and seed 1 can be seen in figure \ref{fig:s2g1s2_engage_summary}. Similar to the first scenario, a lot of the aggressive behavior takes place at the beginning of the simulation among agents of the same group. After the two groups meet, they redirect their anti-social behavior toward members of the other group since the affiliation penalty is then lower.

\begin{figure}[]
	\centering
	\includegraphics[width=0.5\textwidth]{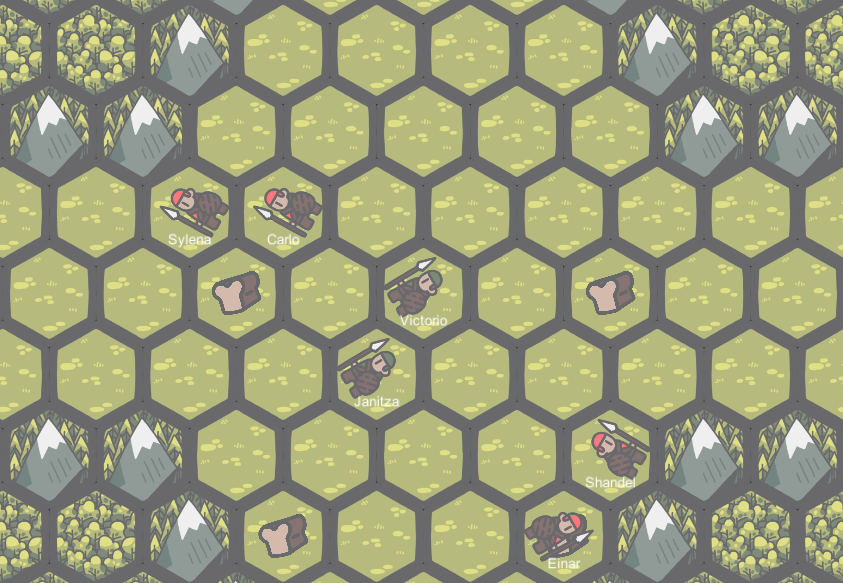}
	\caption{A scene from scenario 2 with group 2 and seed 1 at time step 1000. Three pairs of agents can be seen here executing the standard behavior (social interaction with short food breaks). The pairs from opposing groups do not interfere with one another and therefore cohabit the environment.}
	\label{fig:s2g2s1_cohabit}
\end{figure}

Even though there was occasional fighting between the groups, agents chose most of the time to peacefully cohabit the environment. This can be seen in figure \ref{fig:s2g2s1_cohabit} where two pairs of the red team executed the by now well-known behavioral pattern of social interaction with short food breaks while one pair of the green team is doing the same.
    
Conflict between the two groups usually arises when a single agent of one group with a pronounced need for certainty and for competence meets an agent from the other group. As the engage action plan toward agents with a low social score results in a high certainty reward and a low affiliation punishment, the agent decides to attack.
    
\begin{figure}[]
	\centering
	\includegraphics[width=0.8\textwidth]{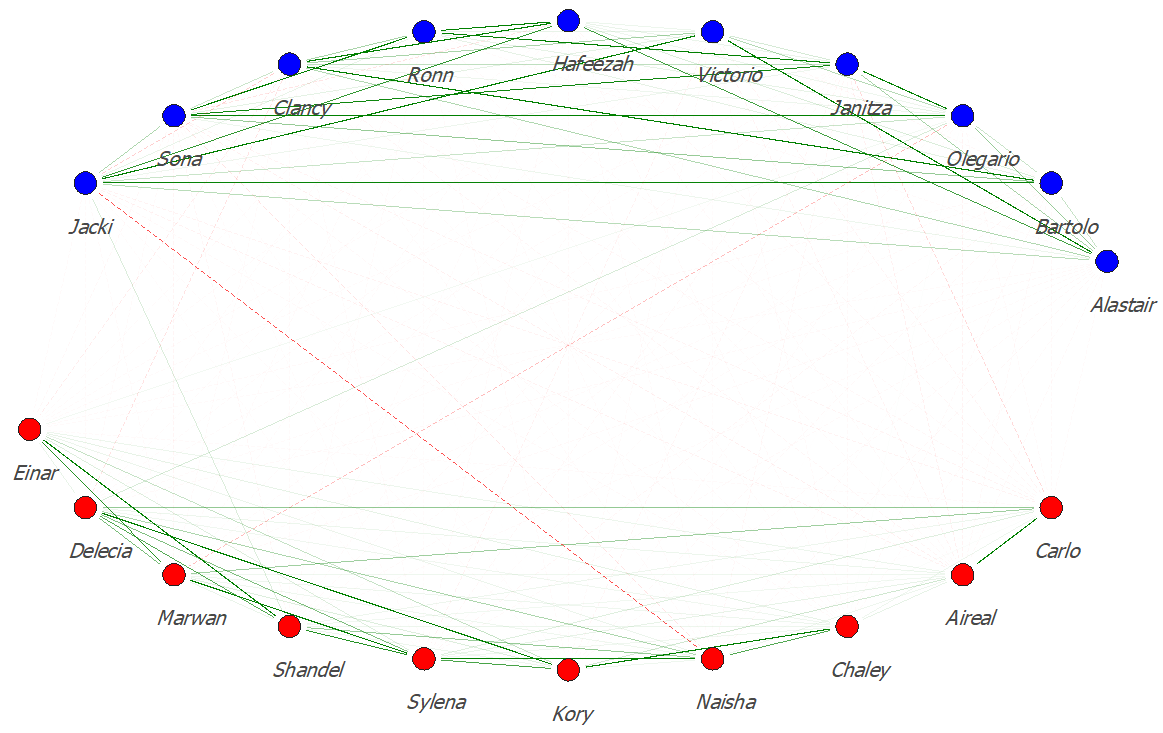}
	\caption{The final social network for scenario 2 with group 2 and seed 1 at time step $t=5000$. The blue nodes represent the members of team 1; the red nodes represent the members of team 2. As we can see, there are almost no strong connections between the two groups of agents. Most of the interaction happens inside the groups. Please note that the social cohesion, i.e., the average social score of group members, is larger compared with the social network of scenario 1 in figure \ref{fig:s1g2s1_social_graph}.}
	\label{fig:s2g2s1_social_network}
\end{figure}

In most runs of this scenario no social interaction across group lines emerged (see figure \ref{fig:s2_information_exchange_summary} in the appendix). Therefore most of the time, the social networks of agents were similar to the one in figure \ref{fig:s2g2s1_social_network}. The relationships between the two teams decayed toward zero because of the lack of inter-group interaction. It is interesting to observe that the number of strong positive relationships increased inside a group. The presence of a second team forced the agents to move around more often to find food and to avoid adversaries. This situation led to a split up of the pairs of agents and forced them to interact with more group members. Through this mechanism, the social cohesion inside a group increased.
    
\begin{figure}[]
	\centering
	\includegraphics[width=0.8\textwidth]{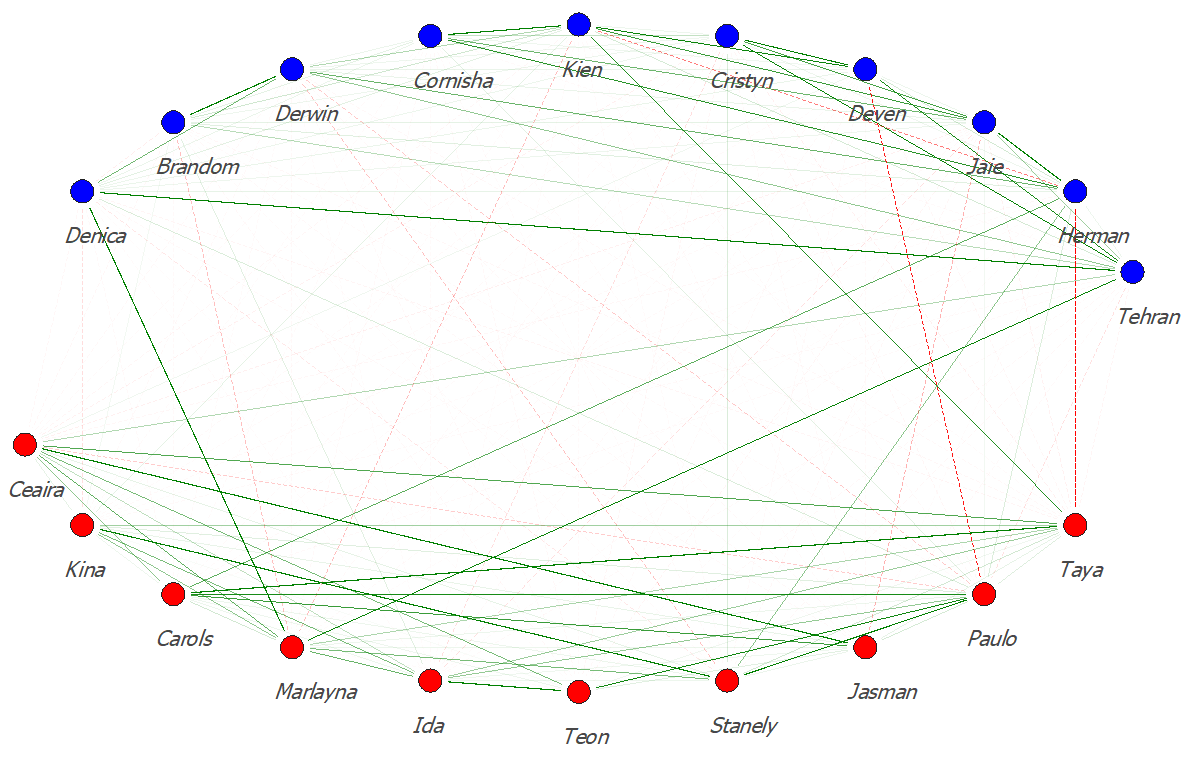}
	\caption{The final social network of scenario 2 with group 1 and seed 1 at time step $t=5000$. Most notable are the strong positive relationships between agents of opposing groups (e.g., between Denica and Marlayna). This shows that social relationships between adversarial groups can emerge even under the harshest of conditions.}
	\label{fig:s2g1s1_social_network}
\end{figure}

In four simulations, social interaction emerged between the two groups. Most prominently, this can be seen in the run with group 1 and seed 1 (see figure \ref{fig:s2g1s1_social_network}). The interaction of two agents from adversarial groups usually happens in a particular case: A lone agent with a low affiliation score meets an agent from the other team. As this single other agent is the only nearby (visible) source of affiliation signals, the agent clings to it. With every interaction, the relationship improves, which then leads to higher affiliation rewards. Through this interaction, the two agents become friends and sometimes begin to execute the standard strategy of social interaction between food breaks.
    
The above-described case does not happen if the affiliation value is low, but the certainty and competence values are even lower. In that case, agents either engage in violent behavior or roam around the environment until they meet a team member with whom social interaction is more rewarding.
    
In conclusion, we argue that the introduction of a second adversarial group mostly fulfills the hypotheses we have made in section \ref{sec:TheScenarios}. The anti-social behavior is redirected mainly against the adversarial group. The existence of the second group also provokes the agents to move around more often and get in contact with more group members. Through this, the social cohesion of the group increases in comparison to scenario 1. What we have not expected was the relatively peaceful coexistence of the two groups. The number of conflicts, though larger compared to scenario 1, was lower than anticipated and limited to only a handful of agents. The social scores between the two groups of agents decayed toward 0. This can be interpreted as a softening of boundaries between the two groups, which might offer an opportunity for the two groups to merge in future.
Surprising was also the rare emergence of relationships between members of different groups. Though this happened only in very specific cases, we did not anticipate such behavior. It shows how it is still possible to overcome social tensions, even in harsh scenarios.

\subsection{Scenario 3: Two Moderately Adversarial Groups}

\begin{figure}[h]
\centering
\includegraphics[width=\textwidth]{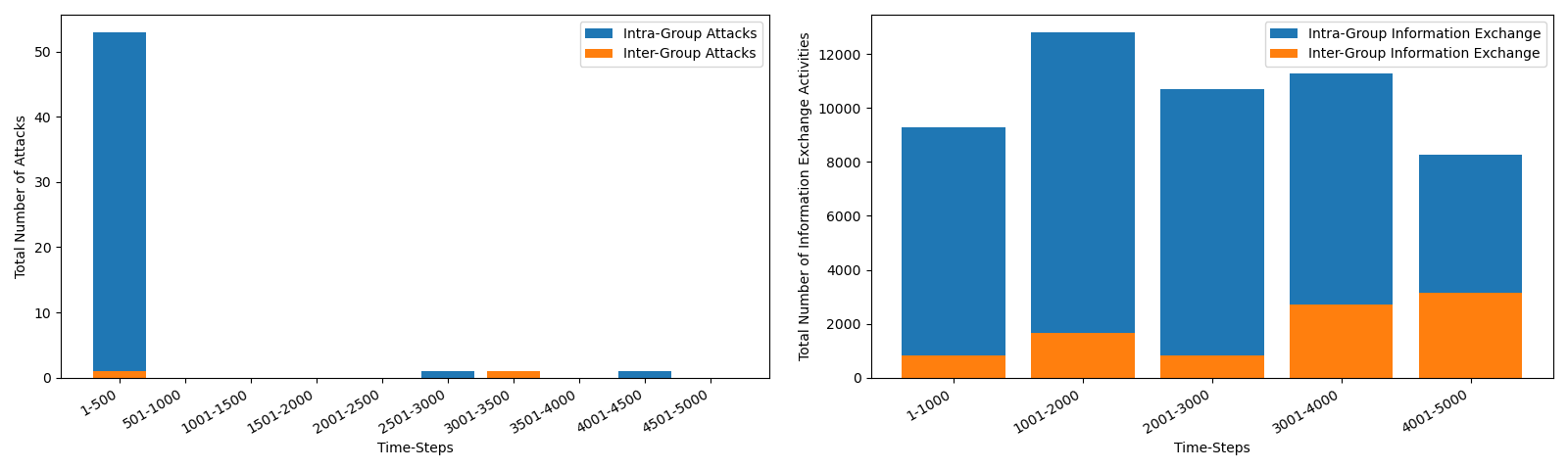}
\caption{Comparison between the counts of engage actions (left) and counts of social interactions (right) split up into intra- and inter-group interactions for scenario 3 with group 1 and seed 2. One can see that the number of attack actions is significantly reduced compared to scenario two, and the number of social interactions between members of the opposing groups has increased. To see all plots for scenario three regarding the engage action, please refer to figure \ref{fig:s3_engage_actionplan_summary} and for social interaction refer to figure \ref{fig:s3_information_exchange_summary} in the appendix.}
\label{fig:s3g1s2_engage_social_comparison}
\end{figure}

In scenario 3, the social interactions between the two groups of agents increased significantly while the number of conflicts between the agents shrunk. A representative comparison between the conflicts and social interactions can be seen in figure \ref{fig:s3g1s2_engage_social_comparison}. Similar to scenario 2, social bonds developed between the two groups of agents. The initially lower negative predisposition toward members of the other group promoted interactions earlier in the simulation that led to overcoming the social divide.

Specific cases where inter-group interactions emerged are very similar to those described in scenario 2: One of the agents had a pronounced need for affiliation while all other needs were relatively satisfied. The social cohesion inside the group still increased due to the existence of the second group.
    
\begin{figure}[h]
	\centering
	\includegraphics[width=0.8\textwidth]{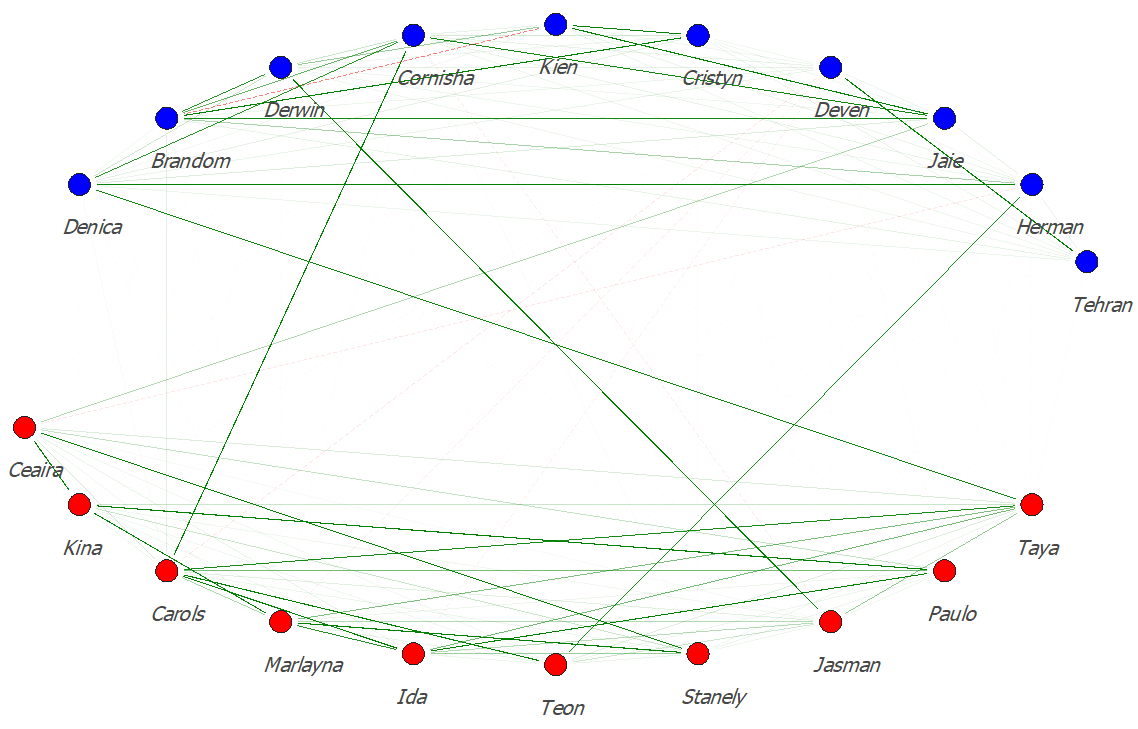}
	\caption{The final social network for scenario 3 with group 1 and seed 2 at time step $t=5000$. The social network shows an increased number of positive relationships between members of the adversarial groups compared to scenario two.}
\label{fig:s3g1s2_social_network}
\end{figure}
    
In our experiments, there was no run without intra-group relations developing. At some point, two agents of the opposing groups got themselves into a situation where they interacted positively. In summary, we can say that the third scenario played out as expected: The presence of a second team forced the agents inside each group to cooperate more often while the social divide between the groups was low enough to promote inter-group exchange. Otherwise, no additional new behaviors were observed.
    
\section{Discussion} \label{sec:ConclusionAndFutureWork}
In this paper, we have analyzed interactions between members of two adversarial groups of PSI agents. The paper shed some light on three questions: How does the introduction of an adversarial group affect the behavior of a society of PSI agents? Does the intra-group social cohesion increase compared to a scenario with only one group of agents? And, finally, do inter-group relationships develop?
    
We have created a simulation using a slightly modified model of the PSI Theory. We have looked at three main scenarios: A scenario with only one group of agents, a scenario with two highly adversarial groups, and a scenario with mildly adversarial groups. Each scenario was run three times for all three pairs of groups. The agents' behavior and the groups' social structure were subsequently analyzed.

In conclusion, we argue that we successfully answered all defined research questions. We have shown that the behavior of the agents changed after the introduction of an adversarial group. The aggressive behavior shifted from intra-group fighting to inter-group fighting. Furthermore, the introduction of the second group of agents forced the members of the same group to move around more often and interact with fellow group. This, in turn, answered our second question: The social cohesion inside a group increased as more positive relationships among members of the same group were maintained. We also successfully found examples of inter-group relationships--these relationships developed in very specific situations which provided incentives for overcoming social boundaries.
    
We are aware that the simulation itself could be tweaked or expanded. One such tweak could be to further calibrate the simulation parameters, such as the need satisfaction values of action plans and the leakage values of the different need tanks. Such calibration could offer new insights into PSI agent behavior. The structure and functionality of the social action plans could also be expanded in a way that would dissuade agents from spending a lot of time with their \enquote{best friends}. This would lead to a shift away from the standard behavior we described earlier, i.e., spending most of the time with one agent friend, save for the occasional food forage periods.
    
Another area to look into is the introduction of reinforcement learning methods into PSI so to create dynamic agents capable of learning and planning using the need satisfaction signals as a reward function. It would also be interesting to see whether similar social structures would arise here as in scenario two and three. Finally, we suggest conducting an analysis of the role of personality in intra- and inter-group behavior. Also, a comparison of agent behavior to animal and human behavior might also prove insightful.

\bibliography{refs}

\newpage

\appendix

\section{Simulation Settings} \label{sec:SimulationSettings}

In this section, we give an overview of the specific values that were used in our simulations. In table \ref{tab:SimulationSettingsGaussianNeedTanks} and table \ref{tab:SimulationSettingsGaussianOther} you can find all $\mu$ and $\sigma$ values that were used for the Gaussian distributions of our simulation. The agent specific values for all set values $set^{need}$ and leakage values $\eta^{need}$ can be found in appendix \ref{sec:AgentPersonalityValues}.

\begin{table}[ht]
    \centering
    \begin{tabular}{|l|c|c|} \hline
        \rowcolor{table_background} Distribution name &  $\mu$  & $\sigma$ \\ \hline
        {\cellcolor{table_background}} $set^{\text{Pain Avoidance}}$  & 0.85 & 0.03 \\ \hline
        {\cellcolor{table_background}} $set^{\text{Energy}}$ & 0.8 & 0.005 \\ \hline
        {\cellcolor{table_background}} $set^{\text{Affiliation}}$ & 0.8 & 0.005 \\ \hline
        {\cellcolor{table_background}} $set^{\text{Certainty}}$ & 0.8 & 0.005 \\ \hline
        {\cellcolor{table_background}} $set^{\text{Competence}}$ & 1 & 0 \\ \hline
        {\cellcolor{table_background}} $\eta^{\text{Energy}}$ & 0.00033 & 0.005 \\ \hline
        {\cellcolor{table_background}} $\eta^{\text{Affiliation}}$ & 0.00033 & 0.005  \\ \hline
        {\cellcolor{table_background}} $\eta^{\text{Certainty}}$ & 0.00033 & 0.005  \\ \hline
        {\cellcolor{table_background}} $\eta^{\text{Competence}}$ & 0.00033 & 0.005  \\ \hline
    \end{tabular}

    \caption{All $\mu$ and $\sigma$ values for the need tanks that define the personality of the agents. The pain avoidance need has a leakage rate of 0 in our simulation and is therefore not represented in this table.}
    \label{tab:SimulationSettingsGaussianNeedTanks}
\end{table}

\begin{table}[ht]
    \centering
    \begin{tabular}{|l|c|c|} \hline
        \rowcolor{table_background} Distribution name & $\mu$ & $\sigma$ \\ \hline
        {\cellcolor{table_background}} Damage Output & $1.8\dot{3}$ & 6.5 \\ \hline
        {\cellcolor{table_background}} Healing Output  & $1.\dot{6}$ & 10 \\ \hline
        {\cellcolor{table_background}} Success Probability Update Alpha & 0.3 & 0.05  \\ \hline
        {\cellcolor{table_background}} S1 New Agent Same Team Score & 0.75 & 0.05  \\ \hline
        {\cellcolor{table_background}} S2 New Agent Same Team Score & 0.75 & 0.05  \\ \hline
        {\cellcolor{table_background}} S3 New Agent Same Team Score & 0.75 & 0.05  \\ \hline
        {\cellcolor{table_background}} S2 New Agent Opposing Team Score & -0.9 & 0.05  \\ \hline
        {\cellcolor{table_background}} S3 New Agent Opposing Team Score & -0.5 & 0.05  \\ \hline
    \end{tabular}
    \caption{The $\mu$ and $\sigma$ values for the damage and healing output, the success probability update alpha (regulating how much influence new experiences have on the predicted success probability of an action plan), and the initial social scores for all scenarios depending on group membership.}
    \label{tab:SimulationSettingsGaussianOther}
\end{table}

The used forgetting rate factors can be found in table \ref{tab:SimulationSettingsLocationForgetRate}. For simplicity reasons, we have decided to keep the forgetting rate for location information the same for all needs.

\begin{table}[ht]
    \centering
    \begin{tabular}{|l|c|} \hline
        \rowcolor{table_background} Forget Rate Type & $\phi^l$  \\ \hline
        {\cellcolor{table_background}} Location Pain Avoidance Positive & 0.9995 \\ \hline
        {\cellcolor{table_background}} Location Pain Avoidance Negative & 0.9985 \\ \hline
        
        {\cellcolor{table_background}} Location Energy Positive & 0.9995 \\ \hline
        {\cellcolor{table_background}} Location Energy Negative & 0.9985 \\ \hline
        
        {\cellcolor{table_background}} Location Affiliation Positive & 0.9995 \\ \hline
        {\cellcolor{table_background}} Location Affiliation Negative & 0.9985 \\ \hline
        
        {\cellcolor{table_background}} Location Certainty Positive & 0.9995 \\ \hline
        {\cellcolor{table_background}} Location Certainty Negative & 0.9985 \\ \hline
        
        {\cellcolor{table_background}} Location Competence Positive & 0.9995 \\ \hline
        {\cellcolor{table_background}} Location Competence Negative & 0.9985 \\ \hline
        
        {\cellcolor{table_background}} Social Positive & 0.9995 \\ \hline
        {\cellcolor{table_background}} Social Negative & 0.9992 \\ \hline
    \end{tabular}
    \caption{The used forgetting rate for location and social information. Each time step $t$ the need satisfaction associations of the location information, and the social score are adapted according to the forgetting rate. For simplicity reasons we have decided to keep the location information forgetting rate the same for all needs.}
    \label{tab:SimulationSettingsLocationForgetRate}
\end{table}

On the exchange of information, an agent's opinion about locations or other agents is regulated using a discount factor. Most importantly, the agent's social opinions are influenced differently depending on whether they know the other agent or not, i.e. if they already have an opinion about the other agent. The new information is then taken into consideration weighted by the discount factors described in table \ref{tab:SimulationSettingsInformationExchange}.

\begin{table}[ht]
    \centering
    \begin{tabular}{|l|c|} \hline
        \rowcolor{table_background} Information Exchange Type & Discount Factor \\ \hline
        
        {\cellcolor{table_background}} Location Information Pain Avoidance & 0.1 \\ \hline
        {\cellcolor{table_background}} Location Information Energy & 0.1 \\ \hline
        {\cellcolor{table_background}} Location Information Affiliation & 0.1 \\ \hline
        {\cellcolor{table_background}} Location Information Certainty & 0.1 \\ \hline
        {\cellcolor{table_background}} Location Information Competence & 0.1 \\ \hline
    
        {\cellcolor{table_background}} Social Information Known-Agent & 0.8 \\ \hline
        {\cellcolor{table_background}} Social Information Unknown-Agent & 0.4 \\ \hline
    \end{tabular}
    \caption{All discount factors that play a role in the exchange of information. On exchange of information the new information is taken into account by the given factors.}
    \label{tab:SimulationSettingsInformationExchange}
\end{table}

\section{Action Plan Settings} \label{sec:ActionPlanValues}

The action plan settings define the rewards / punishments given for a successful or failed execution of an action plan. The concrete value can be found in table \ref{tab:ActionPlanSettings}.
Need satisfaction signals for successful execution of an action plan are marked with $s^{need}$, and unsuccessful execution of action plans are marked with $f^{need}$.
To fit the table on the page we have shortened the needs as follows: Pain avoidance = $p$, Energy = $e$, Affiliation = $a$, Certainty = $ce$, Competence = $co$.

\begin{table}[ht]
    \centering
    \begin{tabular}{|c|c|c|c|c|c|c|c|c|c|c|} \hline
        \rowcolor{table_background} Action-Plan & 
        $s^{p}$ & $s^{e}$ & $s^{a}$ & $s^{ce}$ & $s^{co}$ & $f^{p}$ & $f^{e}$ & $f^{a}$ & $f^{ce}$ & $f^{co}$ \\ \hline
        \cellcolor{table_background} Explore & 0 & 0 & 0 & 0.3 & 0.25 & 0 & 0 & 0 & 0 & 0 \\ \hline
        \cellcolor{table_background} Engage & 0 & 0 & -0.25 & 0.35 & 0.35 & 0 & 0 & -0.25 & -0.3 & -0.4 \\ \hline
        \cellcolor{table_background} Flee & 0.1 & 0 & 0 & 0.1 & 0.05 & -0.15 & 0 & 0 & -0.2 & -0.2 \\ \hline
        \cellcolor{table_background} Search food & 0 & 0.2 & 0 & 0.2 & 0.2 & 0 & 0 & 0 & -0.2 & -0.3 \\ \hline
        \cellcolor{table_background} Call for food & 0 & 0.2 & 0.1 & 0 & 0.05 & 0 & 0 & -0.2 & -0.2 & -0.2 \\ \hline
        \cellcolor{table_background} Give food & 0 & 0 & 0.2 & 0.05 & 0.1 & 0 & 0 & -0.22 & -0.1 & -0.2 \\ \hline
        \cellcolor{table_background} Collect food & 0 & 0 & 0 & 0.1 & 0.1 & 0 & 0 & 0 & -0.2 & -0.2 \\ \hline
        \cellcolor{table_background} General food & 0 & 0.2 & 0 & 0 & 0.25 & 0 & 0 & 0 & -0.25 & -0.3 \\ \hline
        \cellcolor{table_background} Self heal & 0.1 & 0 & 0 & 0 & 0.2 & 0 & 0 & 0 & -0.05 & -0.2 \\ \hline
        \cellcolor{table_background} Request heal & 0.1 & 0 & 0.2 & 0 & 0.05 & 0 & 0 & -0.2 & -0.05 & -0.2 \\ \hline
        \cellcolor{table_background} Go heal & 0 & 0 & 0.2 & 0.05 & 0.1 & 0 & 0 & -0.2 & -0.1 & -0.2 \\ \hline
        \cellcolor{table_background} Exchange info & 0 & 0 & 0.2 & 0.05 & 0.05 & 0 & 0 & -0.2 & -0.15 & -0.2 \\ \hline
    \end{tabular}
    \caption{Table of need satisfaction signals for successful behavior $s$ and unsuccessful behavior $f$.}
    \label{tab:ActionPlanSettings}
\end{table}

One might notice that not all action plans are mentioned in table \ref{tab:ActionPlanSettings}. This is the case as some action plans share their need satisfaction signals with others:

\begin{itemize}
    \item \enquote{Exchange info} gives the need satisfaction signal for the exchange of social and location information.
    \item \enquote{General food} gives the need satisfaction signals for searching and eating food from a known food cluster and eating food from storage.
\end{itemize}

Some need satisfaction signals depend on the relationship between two agents. This is especially the case for affiliation signals and certainty signals. Therefore some action plans adapt their rewards and punishments using functions that depend on the social score. 

In many cases, the affiliation signal is adapted by a factor $\lambda_{a}$ according to the social score $s$ using the function:

\begin{equation*}
    \lambda_{a} = s + 1
\end{equation*}

It follows that $\lambda_{a} \in [0; 2]$. Therefore, the better the relationship is between the agents, the greater the affiliation reward or punishment. This encourages social behavior and discourages anti-social behavior towards friendly agents. The adaption function is used in all social action plans, i.e., exchange of information, sharing food, or healing another agent, as well as in the engage action plan.

As defined in section \ref{sec:NeedStructure} certainty is the need to feel confident about the beliefs the agent has about the world. Some action plans have the property of reinforcing the agent's beliefs about the environment or destroying them. In our simulation, this is the case for the engage action plan: If an agent is attacked or it attacks an agent, they have a negative relationship. However, their certainty will increase as their negative beliefs are verified. If they are attacked or attack an agent, they have a positive relationship, with their certainty should decrease. We have modeled this using a factor $\lambda_{ce}$ that is dependent on the social score $s$:

\begin{equation*}
    \lambda_{ce} = -s^{3}
\end{equation*}

Using this weight factor the agent gets a positive certainty signal for $s < 0$ and a negative certainty signal for $s > 0$.

The final need satisfaction signal is then calculated as the product of the need satisfaction signal defined in table \ref{tab:ActionPlanSettings} and the weight factor $\lambda$.
 
These functions were defined using an intuitive approach and are not directly backed by psychological research. Verifying these functions and adapting the weight factors is left to future work.

\newpage

\section{Agent Personality} \label{sec:AgentPersonalityValues}
As described in section \ref{sec:NeedStructure}, each agent has its own personality emerging from its specific set values $set^{need}$ and leakage values $\eta^{need}$ of their need tanks.
The specific set values for each agent of each team can be found in table \ref{tab:SetValuesSeed1}, \ref{tab:SetValuesSeed2}, and \ref{tab:SetValuesSeed3}. The values for the leakage can be found in the tables \ref{tab:LeakageSeed1}, \ref{tab:LeakageSeed2}, and \ref{tab:LeakageSeed3}. The specific teams were generated using the seeds 1, 2, and 3.

As the set value for Competence is 1, and the leakage value for Pain Avoidance is 0 for all generated personalities, they will not be mentioned in the tables below.

\begin{landscape}
     \begin{table}[ht]
    	\centering
    	\begin{tabular}{|c|c|c|c|c|}
    		\hline
    		\rowcolor{table_background} Name       & $set^{PainAvoidance}$ & $set^{Energy}$ & $set^{Affiliation}$ & $set^{Certainty}$  \\
    		\hline 
    		{\cellcolor{table_background}} Brandom & 0.832322710251051 & 0.792954903009047 & 0.800698923499392 & 0.79635015154248 \\
    
    		\hline
    		{\cellcolor{table_background}} Cornisha & 0.856498127147759 & 0.798857921703419 & 0.800891525368565 & 0.807310926345435 \\
    
    		\hline
    		{\cellcolor{table_background}} Cristyn & 0.831100524696322 & 0.796563911665623 & 0.809200022236316 & 0.796765568495608 \\
    
    		\hline
    		{\cellcolor{table_background}} Denica & 0.793050211001376 & 0.797831122688682 & 0.808932545301169 & 0.80344894206837 \\
    
    		\hline
    		{\cellcolor{table_background}} Derwin & 0.91101570476529 & 0.797474283435326 & 0.8066887096708 & 0.800946344989395 \\
    
    		\hline
    		{\cellcolor{table_background}} Deven & 0.838068606546197 & 0.805718655012395 & 0.799721555785536 & 0.792666793718805 \\
    
    		\hline
    		{\cellcolor{table_background}} Herman & 0.870946021665327 & 0.801094975803025 & 0.798376641905167 & 0.794785425672169 \\
    
    		\hline
    		{\cellcolor{table_background}} Jaie & 0.86599910645479 & 0.795886276690885 & 0.798870127772047 & 0.806060822180624 \\
    
    		\hline
    		{\cellcolor{table_background}} Kien & 0.860825896735131 & 0.796193652159422 & 0.793217745686873 & 0.790093328980956 \\
    
    		\hline
    		{\cellcolor{table_background}} Tehran & 0.791591828039676 & 0.797750892535971 & 0.808123956817756 & 0.794490945045974 \\
    
    		\hline
    	\end{tabular}
    
    	\begin{tabular}{|c|c|c|c|c|}
    		\hline
    		\rowcolor{table_background} Name        & $set^{PainAvoidance}$ & $set^{Energy}$ & $set^{Affiliation}$ & $set^{Certainty}$  \\
    		\hline 
    		{\cellcolor{table_background}} Carols & 0.850985342280945 &  0.795826540279336 & 0.796339358823885 & 0.79856189172767 \\
    
    		\hline
    		{\cellcolor{table_background}} Ceaira & 0.850107741818634 &  0.805745045284645 & 0.800984060347145 & 0.796777907790234 \\
    
    		\hline
    		{\cellcolor{table_background}} Ida & 0.895180354743987 &  0.799532992558879 & 0.795944085035451 & 0.799949416635265 \\
    
    		\hline
    		{\cellcolor{table_background}} Jasman & 0.885828854392381 &  0.797966503870054 & 0.797594004841705 & 0.798744184153471 \\
    
    		\hline
    		{\cellcolor{table_background}} Kina & 0.853960478541093 &  0.806012170357492 & 0.806808961068263 & 0.802481795780453 \\
    
    		\hline
    		{\cellcolor{table_background}} Marlayna & 0.840506831676294 &  0.795538199346907 & 0.801092467978905 & 0.802168082728738 \\
    
    		\hline
    		{\cellcolor{table_background}} Paulo & 0.883117183855402 &  0.804811080883959 & 0.798309853982689 & 0.811686654605819 \\
    
    		\hline
    		{\cellcolor{table_background}} Stanely & 0.855720353080281 &  0.802309906816761 & 0.810434925731226 & 0.800115380739747 \\
    
    		\hline
    		{\cellcolor{table_background}} Taya & 0.86155734290407 &  0.794659004733448 & 0.792772651896183 & 0.792642268612411 \\
    
    		\hline
    		{\cellcolor{table_background}} Teon & 0.833840826495436 &  0.792112715090178 & 0.79853501515201 & 0.800335412980245 \\
    
    		\hline
    	\end{tabular}
    	\caption{Set-values for team 1 (top) and team 2 (bottom) for seed 1}
    	\label{tab:SetValuesSeed1}
    \end{table}
    
    \begin{table}[ht]
    	\centering
    	\begin{tabular}{|c|c|c|c|c|}
    		\hline
    		\rowcolor{table_background} Name       & $\eta^{Energy}$ & $\eta^{Affiliation}$ & $\eta^{Certainty}$ & $\eta^{Competence}$  \\
    		\hline 
    		{\cellcolor{table_background}} Brandom & 0.00513131392255188 & 0.00459934507003216 & 0.00494593232012495 & 0.00449886894180903 \\
    
    		\hline
    		{\cellcolor{table_background}} Cornisha & 0.00504106784992906 & 0.00492662229707347 & 0.00575408027562202 & 0.00511679276409423 \\
    
    		\hline
    		{\cellcolor{table_background}} Cristyn & 0.00526039934001146 & 0.00469525040634634 & 0.0045107889070089 & 0.00464222987657288 \\
    
    		\hline
    		{\cellcolor{table_background}} Denica & 0.00488802358521213 & 0.00469271668446248 & 0.0046073922759557 & 0.00490158980941015 \\
    
    		\hline
    		{\cellcolor{table_background}} Derwin & 0.00451827261434796 & 0.00498592660720679 & 0.00445752088320257 & 0.00504084484132216 \\
    
    		\hline
    		{\cellcolor{table_background}} Deven & 0.00469997681002749 & 0.00488313819626353 & 0.00514698256408433 & 0.00555297923326447 \\
    
    		\hline
    		{\cellcolor{table_background}} Herman & 0.00523322112902874 & 0.00542665256228831 & 0.00506181750650757 & 0.00524011100641141 \\
    
    		\hline
    		{\cellcolor{table_background}} Jaie & 0.00432559742055996 & 0.0055605519527752 & 0.00500502916780431 & 0.00529130103726344 \\
    
    		\hline
    		{\cellcolor{table_background}} Kien & 0.00479090791950936 & 0.0046738765220001 & 0.00474016490931056 & 0.00482586506528426 \\
    
    		\hline
    		{\cellcolor{table_background}} Tehran & 0.00541438498501485 & 0.0054264135425439 & 0.00547314820852489 & 0.00513104453458033 \\
    
    		\hline
    	\end{tabular}
    
    	\begin{tabular}{|c|c|c|c|c|}
    		\hline
    		\rowcolor{table_background} Name        & $\eta^{Energy}$ & $\eta^{Affiliation}$ & $\eta^{Certainty}$ & $\eta^{Competence}$  \\
    		\hline 
    		{\cellcolor{table_background}} Carols & 0.00484773777330993 & 0.00511349577696493 & 0.00427520237511847 & 0.00511085745561829 \\
    
    		\hline
    		{\cellcolor{table_background}} Ceaira & 0.00496202942882203 & 0.00452811574381595 & 0.00548720246049735 & 0.00440840679139356 \\
    
    		\hline
    		{\cellcolor{table_background}} Ida & 0.00461041115137068 & 0.00491362487813937 & 0.00431521992423047 & 0.00501723433195968 \\
    
    		\hline
    		{\cellcolor{table_background}} Jasman & 0.00498662220432355 & 0.00474467973466491 & 0.0052927036220236 & 0.00502027836119064 \\
    
    		\hline
    		{\cellcolor{table_background}} Kina & 0.00450612244064935 & 0.00515387080648576 & 0.00486853596303127 & 0.00525520779661425 \\
    
    		\hline
    		{\cellcolor{table_background}} Marlayna & 0.00503681551864251 & 0.00510343091366898 & 0.00476480610807922 & 0.00513468004269447 \\
    
    		\hline
    		{\cellcolor{table_background}} Paulo & 0.00509178097462441 & 0.00491471539186201 & 0.00469687820781739 & 0.00508371560206849 \\
    
    		\hline
    		{\cellcolor{table_background}} Stanely & 0.00485312232984306 & 0.00478001760980506 & 0.00490281543833036 & 0.00501540382845146 \\
    
    		\hline
    		{\cellcolor{table_background}} Taya & 0.0046235107144188 & 0.00460050333508733 & 0.00548305712488281 & 0.00549043231155397 \\
    
    		\hline
    		{\cellcolor{table_background}} Teon & 0.00497959999466033 & 0.00482668564329631 & 0.00478208842424309 & 0.00448396457698363 \\
    
    		\hline
    	\end{tabular}
    	\caption{Leakage values for team 1 (top) and team 2 (bottom) for seed 1}
    	\label{tab:LeakageSeed1}
    \end{table}
    
    \begin{table}[ht]
    	\centering
    	\begin{tabular}{|c|c|c|c|c|}
    		\hline
    		\rowcolor{table_background} Name       & $set^{PainAvoidance}$ & $set^{Energy}$ & $set^{Affiliation}$ & $set^{Certainty}$  \\
    		\hline 
    		{\cellcolor{table_background}} Alastair & 0.855583475863848 & 0.796237965927645 & 0.805797526057067 & 0.795959823079966 \\
    
    		\hline
    		{\cellcolor{table_background}} Bartolo & 0.86367830194419 & 0.804482637483858 & 0.802368053301973 & 0.797451397614642 \\
    
    		\hline
    		{\cellcolor{table_background}} Clancy & 0.831824670104804 & 0.808078605644282 & 0.79829144945311 & 0.797248782052195 \\
    
    		\hline
    		{\cellcolor{table_background}} Hafeezah & 0.874596117908771 & 0.8051801031229 & 0.791568673903932 & 0.801168669621973 \\
    
    		\hline
    		{\cellcolor{table_background}} Jacki & 0.827210727455071 & 0.808904990702906 & 0.795458872543204 & 0.804421687703922 \\
    
    		\hline
    		{\cellcolor{table_background}} Janitza & 0.81473035102438 & 0.803790792106462 & 0.801073548370917 & 0.803977463393151 \\
    
    		\hline
    		{\cellcolor{table_background}} Olegario & 0.859426119502748 & 0.809126465009003 & 0.79177179585641 & 0.79865036949277 \\
    
    		\hline
    		{\cellcolor{table_background}} Ronn & 0.798182355283083 & 0.797092013848232 & 0.808515964655234 & 0.80967559460009 \\
    
    		\hline
    		{\cellcolor{table_background}} Sona & 0.873667548278202 & 0.794085011395871 & 0.794457248638175 & 0.794856679044116 \\
    
    		\hline
    		{\cellcolor{table_background}} Victorio & 0.883247351791622 & 0.791869198290374 & 0.810982075932905 & 0.80395318005968 \\
    
    		\hline
    	\end{tabular}
    
    	\begin{tabular}{|c|c|c|c|c|}
    		\hline
    		\rowcolor{table_background} Name        & $set^{PainAvoidance}$ & $set^{Energy}$ & $set^{Affiliation}$ & $set^{Certainty}$  \\
    		\hline 
    		{\cellcolor{table_background}} Aireal & 0.893858617573515 &  0.799142562599104 & 0.794320187964594 & 0.798225964119081 \\
    
    		\hline
    		{\cellcolor{table_background}} Carlo & 0.878797987212938 &  0.79939603722884 & 0.798681338418294 & 0.796697652689187 \\
    
    		\hline
    		{\cellcolor{table_background}} Chaley & 0.851621451542552 &  0.806127423134215 & 0.794207939030385 & 0.803097677730621 \\
    
    		\hline
    		{\cellcolor{table_background}} Delecia & 0.837463926522748 &  0.801113618625468 & 0.799810409295522 & 0.799817479351591 \\
    
    		\hline
    		{\cellcolor{table_background}} Einar & 0.88805207022194 &  0.802082890021173 & 0.802886867026035 & 0.793336726870717 \\
    
    		\hline
    		{\cellcolor{table_background}} Kory & 0.875238392783338 &  0.807855591462525 & 0.799899987510027 & 0.793868597682221 \\
    
    		\hline
    		{\cellcolor{table_background}} Marwan & 0.886979116227993 &  0.80791362617575 & 0.800930910210786 & 0.812696449180493 \\
    
    		\hline
    		{\cellcolor{table_background}} Naisha & 0.877411775919937 &  0.801642321863907 & 0.796358528619439 & 0.794795334532733 \\
    
    		\hline
    		{\cellcolor{table_background}} Shandel & 0.855284898745799 &  0.796034249655745 & 0.810367731468783 & 0.803302010222193 \\
    
    		\hline
    		{\cellcolor{table_background}} Sylena & 0.862456134293174 &  0.797059401143845 & 0.806021729262493 & 0.800849685111217 \\
    
    		\hline
    	\end{tabular}
    	\caption{Set-values for team 1 (top) and team 2 (bottom) for seed 2}
    	\label{tab:SetValuesSeed2}
    \end{table}

    \begin{table}[ht]
    	\centering
    	\begin{tabular}{|c|c|c|c|c|}
    		\hline
    		\rowcolor{table_background} Name       & $\eta^{Energy}$ & $\eta^{Affiliation}$ & $\eta^{Certainty}$ & $\eta^{Competence}$  \\
    		\hline 
    		{\cellcolor{table_background}} Alastair & 0.00508149915856879 & 0.00513846945699468 & 0.00508607163058062 & 0.00530175071008716 \\
    
    		\hline
    		{\cellcolor{table_background}} Bartolo & 0.00544505050537054 & 0.0047879491087546 & 0.00498181527532912 & 0.00575105321547221 \\
    
    		\hline
    		{\cellcolor{table_background}} Clancy & 0.00484765690894009 & 0.00516323484527528 & 0.00493594003562197 & 0.0049611062188215 \\
    
    		\hline
    		{\cellcolor{table_background}} Hafeezah & 0.00511187443042735 & 0.00479783355494733 & 0.00464043307452539 & 0.00521120568125439 \\
    
    		\hline
    		{\cellcolor{table_background}} Jacki & 0.00498993198253587 & 0.00560935380567736 & 0.00509454311370784 & 0.00499310882458728 \\
    
    		\hline
    		{\cellcolor{table_background}} Janitza & 0.00454262131188636 & 0.0050597819471528 & 0.00493350742167228 & 0.00515690627686317 \\
    
    		\hline
    		{\cellcolor{table_background}} Olegario & 0.00462062911995 & 0.00480883604323403 & 0.00496947388248217 & 0.00492563132213143 \\
    
    		\hline
    		{\cellcolor{table_background}} Ronn & 0.00453738322679906 & 0.00520490461026432 & 0.0048107056469371 & 0.00502551700279798 \\
    
    		\hline
    		{\cellcolor{table_background}} Sona & 0.00544267160542367 & 0.00531022132727168 & 0.00442836658231984 & 0.00499297246814949 \\
    
    		\hline
    		{\cellcolor{table_background}} Victorio & 0.00481008496244654 & 0.00526977615207422 & 0.0054575908736804 & 0.00505508498706451 \\
    
    		\hline
    	\end{tabular}
    
    	\begin{tabular}{|c|c|c|c|c|}
    		\hline
    		\rowcolor{table_background} Name        & $\eta^{Energy}$ & $\eta^{Affiliation}$ & $\eta^{Certainty}$ & $\eta^{Competence}$  \\
    		\hline 
    		{\cellcolor{table_background}} Aireal & 0.00490704377336852 & 0.00520252205159901 & 0.00515614696673266 & 0.00524954707431498 \\
    
    		\hline
    		{\cellcolor{table_background}} Carlo & 0.00471574065400298 & 0.00473912620543458 & 0.00472595388776217 & 0.0052665974664884 \\
    
    		\hline
    		{\cellcolor{table_background}} Chaley & 0.00626392970700487 & 0.00458995551190567 & 0.00521851429226658 & 0.00471309499946593 \\
    
    		\hline
    		{\cellcolor{table_background}} Delecia & 0.00482388923925784 & 0.00454812420495269 & 0.00532216987072333 & 0.005054657563698 \\
    
    		\hline
    		{\cellcolor{table_background}} Einar & 0.00469582389916539 & 0.00555462239923383 & 0.00544056988803851 & 0.00480324273871818 \\
    
    		\hline
    		{\cellcolor{table_background}} Kory & 0.0053155171373737 & 0.00507712295578938 & 0.00516839236983244 & 0.00496847591276298 \\
    
    		\hline
    		{\cellcolor{table_background}} Marwan & 0.00547310929748819 & 0.00500201791116977 & 0.00496641139335839 & 0.00508451041305304 \\
    
    		\hline
    		{\cellcolor{table_background}} Naisha & 0.00482577138492016 & 0.00525764425188747 & 0.00475540376429377 & 0.00495461192032049 \\
    
    		\hline
    		{\cellcolor{table_background}} Shandel & 0.00509620093983623 & 0.00444772766722484 & 0.00572950536360631 & 0.0047368195730885 \\
    
    		\hline
    		{\cellcolor{table_background}} Sylena & 0.0049687884951016 & 0.00535661160300512 & 0.0049955770609861 & 0.00437644186576806 \\
    
    		\hline
    	\end{tabular}
    	\caption{Leakage values for team 1 (top) and team 2 (bottom) for seed 2}
    	\label{tab:LeakageSeed2}
    \end{table}
    
    \begin{table}[ht]
    	\centering
    	\begin{tabular}{|c|c|c|c|c|}
    		\hline
    		\rowcolor{table_background} Name       & $set^{PainAvoidance}$ & $set^{Energy}$ & $set^{Affiliation}$ & $set^{Certainty}$  \\
    		\hline 
    		{\cellcolor{table_background}} Ankur & 0.848489268381742 & 0.800505386163384 & 0.80051577101058 & 0.799065439962204 \\
    
    		\hline
    		{\cellcolor{table_background}} Britten & 0.826843735216339 & 0.802793212428261 & 0.796730708711385 & 0.807836202280559 \\
    
    		\hline
    		{\cellcolor{table_background}} Broc & 0.868043665073195 & 0.806454310898379 & 0.802837852947038 & 0.78664543382539 \\
    
    		\hline
    		{\cellcolor{table_background}} Carles & 0.843306884157594 & 0.814874673098035 & 0.797546806030243 & 0.806348529829497 \\
    
    		\hline
    		{\cellcolor{table_background}} Grayce & 0.861687624951192 & 0.805943977159319 & 0.803122691176525 & 0.794625960301517 \\
    
    		\hline
    		{\cellcolor{table_background}} Jahmal & 0.847205693392775 & 0.802958753003385 & 0.794873920114453 & 0.803783460492238 \\
    
    		\hline
    		{\cellcolor{table_background}} Juvencio & 0.858333885810621 & 0.796489491958884 & 0.797826141880197 & 0.799347334740789 \\
    
    		\hline
    		{\cellcolor{table_background}} Lasha & 0.821775906538399 & 0.797321022943005 & 0.797496551330447 & 0.799579476190369 \\
    
    		\hline
    		{\cellcolor{table_background}} Raine & 0.835068696348766 & 0.799344486357685 & 0.799572390801115 & 0.795311453583102 \\
    
    		\hline
    		{\cellcolor{table_background}} Teddrick & 0.909340907509032 & 0.791629823749638 & 0.795599592215778 & 0.803786919786626 \\
    
    		\hline
    	\end{tabular}
    
    	\begin{tabular}{|c|c|c|c|c|}
    		\hline
    		\rowcolor{table_background} Name        & $set^{PainAvoidance}$ & $set^{Energy}$ & $set^{Affiliation}$ & $set^{Certainty}$  \\
    		\hline 
    		{\cellcolor{table_background}} Agapito & 0.845184640484786 &  0.797867903061935 & 0.800891090001254 & 0.797731349995964 \\
    
    		\hline
    		{\cellcolor{table_background}} Avigdor & 0.900877972307907 &  0.804090596705341 & 0.800674293632414 & 0.79291554635737 \\
    
    		\hline
    		{\cellcolor{table_background}} Christella & 0.795455801568353 &  0.805426684884301 & 0.794320769090438 & 0.803206833467688 \\
    
    		\hline
    		{\cellcolor{table_background}} Doris & 0.849904565240084 &  0.796950441430324 & 0.794795573218907 & 0.796732799349214 \\
    
    		\hline
    		{\cellcolor{table_background}} Fatima & 0.892393459535486 &  0.795342374389936 & 0.79577089315818 & 0.803807821850891 \\
    
    		\hline
    		{\cellcolor{table_background}} Labrandon & 0.885941184407295 &  0.800650038775201 & 0.796066621885089 & 0.80569163985047 \\
    
    		\hline
    		{\cellcolor{table_background}} Lynden & 0.835843688485114 &  0.805422502179591 & 0.800032348942704 & 0.793047897249472 \\
    
    		\hline
    		{\cellcolor{table_background}} Meegan & 0.853017740571288 &  0.798616839083192 & 0.792088334719021 & 0.795499146785824 \\
    
    		\hline
    		{\cellcolor{table_background}} Rashun & 0.839436978344675 &  0.80825820057485 & 0.794360281814363 & 0.800593549162416 \\
    
    		\hline
    		{\cellcolor{table_background}} Webster & 0.866552591311768 &  0.801028288418525 & 0.795488822805204 & 0.792873916553349 \\
    
    		\hline
    	\end{tabular}
    	\caption{Set-values for team 1 (top) and team 2 (bottom) for seed 3}
    	\label{tab:SetValuesSeed3}
    \end{table}

    \begin{table}[ht]
    	\centering
    	\begin{tabular}{|c|c|c|c|c|}
    		\hline
    		\rowcolor{table_background} Name       & $\eta^{Energy}$ & $\eta^{Affiliation}$ & $\eta^{Certainty}$ & $\eta^{Competence}$  \\
    		\hline 
    		{\cellcolor{table_background}} Ankur & 0.00529416583496529 & 0.00591172742695811 & 0.00503991663295102 & 0.00548239408664209 \\
    
    		\hline
    		{\cellcolor{table_background}} Britten & 0.00489909738553782 & 0.00497958275456091 & 0.0049282655818374 & 0.00476719014002268 \\
    
    		\hline
    		{\cellcolor{table_background}} Broc & 0.00461135510268119 & 0.00492102588319946 & 0.00455061076491119 & 0.00513009322607511 \\
    
    		\hline
    		{\cellcolor{table_background}} Carles & 0.0051417478140509 & 0.00461769624842767 & 0.00487933661248768 & 0.00499764944416151 \\
    
    		\hline
    		{\cellcolor{table_background}} Grayce & 0.00573302088951477 & 0.00509622955488778 & 0.00551947651127312 & 0.00444803326278986 \\
    
    		\hline
    		{\cellcolor{table_background}} Jahmal & 0.0050152487553554 & 0.00472092324772384 & 0.00490178331883663 & 0.00517561330104557 \\
    
    		\hline
    		{\cellcolor{table_background}} Juvencio & 0.00557914814125437 & 0.00471678927885474 & 0.00503338914807129 & 0.00446459001513936 \\
    
    		\hline
    		{\cellcolor{table_background}} Lasha & 0.00485407199027455 & 0.00525998840493339 & 0.00465946913990278 & 0.00483667264819984 \\
    
    		\hline
    		{\cellcolor{table_background}} Raine & 0.00529608116691271 & 0.00482491898701301 & 0.00580729686538253 & 0.00521987615980999 \\
    
    		\hline
    		{\cellcolor{table_background}} Teddrick & 0.00486329436291201 & 0.00470148000369855 & 0.00460294468744031 & 0.00545132375810979 \\
    
    		\hline
    	\end{tabular}
    
    	\begin{tabular}{|c|c|c|c|c|}
    		\hline
    		\rowcolor{table_background} Name        & $\eta^{Energy}$ & $\eta^{Affiliation}$ & $\eta^{Certainty}$ & $\eta^{Competence}$  \\
    		\hline 
    		{\cellcolor{table_background}} Agapito & 0.00440095664733108 & 0.00506304671621021 & 0.005335595910605 & 0.00554990807024724 \\
    
    		\hline
    		{\cellcolor{table_background}} Avigdor & 0.00419325239626519 & 0.00447374970022898 & 0.00505966533624873 & 0.00485370193342157 \\
    
    		\hline
    		{\cellcolor{table_background}} Christella & 0.00508710728944094 & 0.00532312894716248 & 0.00512659531824088 & 0.00459683685607355 \\
    
    		\hline
    		{\cellcolor{table_background}} Doris & 0.00524337858297393 & 0.00507028565927241 & 0.00523118132111844 & 0.00513513918610015 \\
    
    		\hline
    		{\cellcolor{table_background}} Fatima & 0.00471002902991589 & 0.00467487689169888 & 0.00463408038329096 & 0.005353480278037 \\
    
    		\hline
    		{\cellcolor{table_background}} Labrandon & 0.00459324096158953 & 0.00502420314268286 & 0.00448338101216563 & 0.00495577281674202 \\
    
    		\hline
    		{\cellcolor{table_background}} Lynden & 0.00562995267338625 & 0.00463541714811359 & 0.00508851000739389 & 0.0053634801869334 \\
    
    		\hline
    		{\cellcolor{table_background}} Meegan & 0.00468709461070324 & 0.00443992604394482 & 0.00513029149200028 & 0.00526574049831105 \\
    
    		\hline
    		{\cellcolor{table_background}} Rashun & 0.00447707436327303 & 0.00502104653302313 & 0.00483750484534214 & 0.00496246214656275 \\
    
    		\hline
    		{\cellcolor{table_background}} Webster & 0.00497976453550607 & 0.00519334303944919 & 0.00458348772660402 & 0.00497036439939732 \\
    
    		\hline
    	\end{tabular}
    	\caption{Leakage values for team 1 (top) and team 2 (bottom) for seed 3}
    	\label{tab:LeakageSeed3}
    \end{table}

\end{landscape}

\newpage

\section{Figures}
    In this section we find all figures that were generated for this project. Figure \ref{fig:s1_engage_actionplan_summary} to \ref{fig:s3_engage_actionplan_summary} give the counts of all engage actions differentiated by intra- and inter-group attacks summarized in 500 time-step bins. Figure \ref{fig:s1_social_action_summary} to \ref{fig:s3_social_action_summary} summarize the counts of intra- and inter-group social behavior. Figure \ref{fig:s2_information_exchange_summary} and \ref{fig:s3_information_exchange_summary} give the counts of intra- and inter-group information exchange actions.
 
\begin{landscape}

    \begin{figure}
        \centering
        \includegraphics[height=0.85\textwidth]{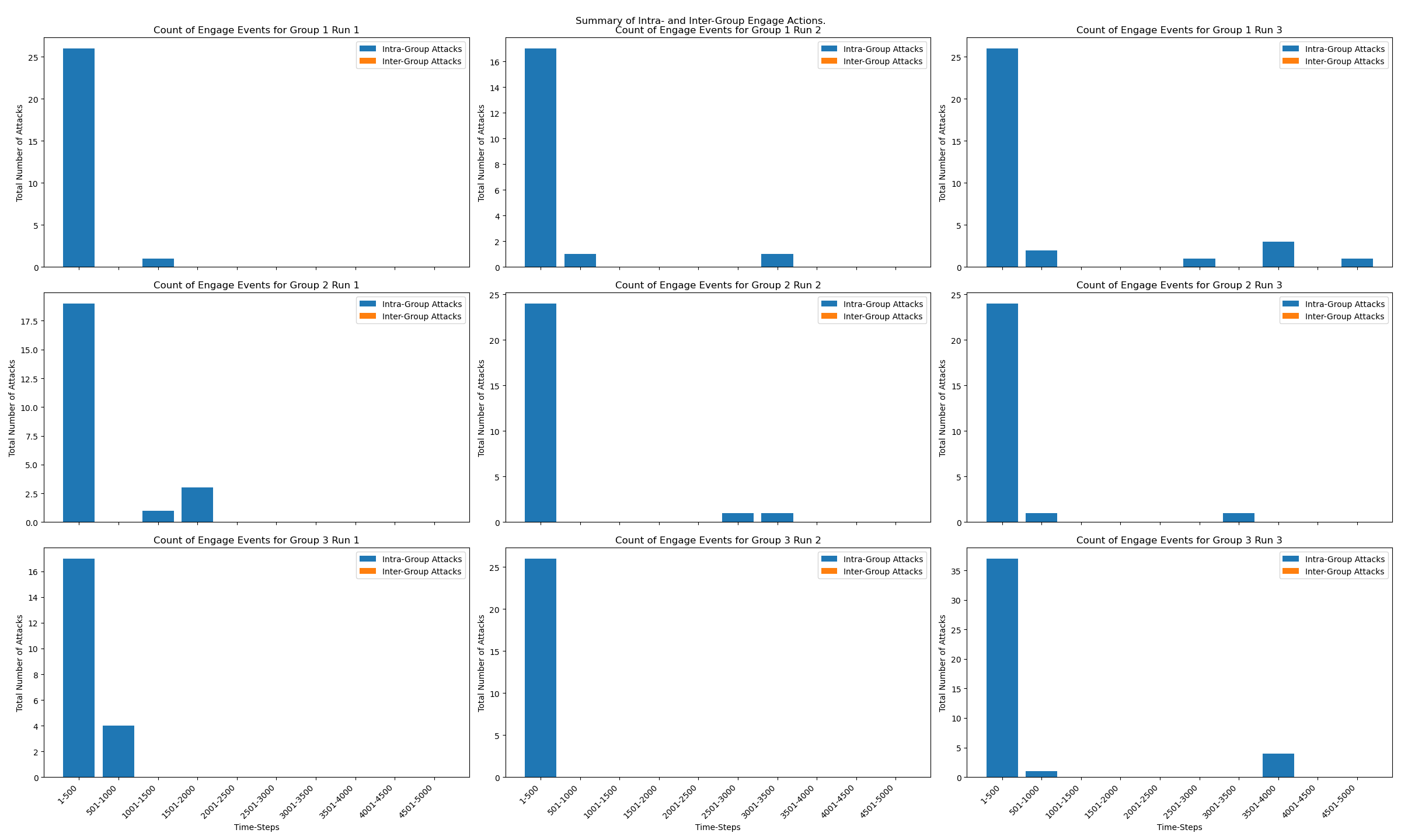}
        \caption{Summary over all engage actions that were taken in scenario 1.}
        \label{fig:s1_engage_actionplan_summary}
    \end{figure}
    
    \begin{figure}
        \centering
        \includegraphics[height=0.85\textwidth]{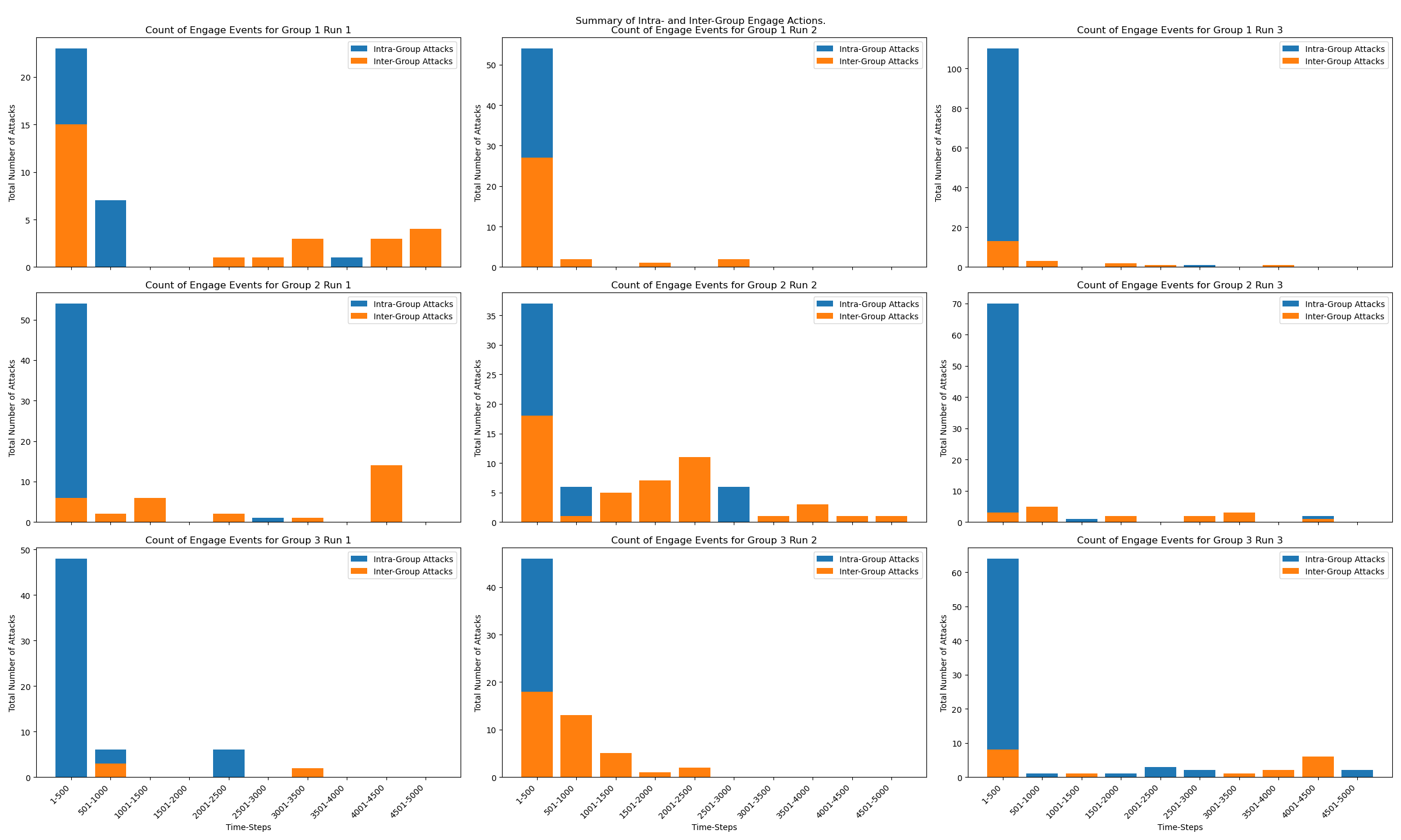}
        \caption{Summary over all engage actions that were taken in scenario 2. The orange part of the bar shows the number of engage actions that were executed between two agents of different groups. The blue part of the bar shows the number of engage actions executed group internally.}
        \label{fig:s2_engage_actionplan_summary}
    \end{figure}
    
    \begin{figure}
        \centering
        \includegraphics[height=0.85\textwidth]{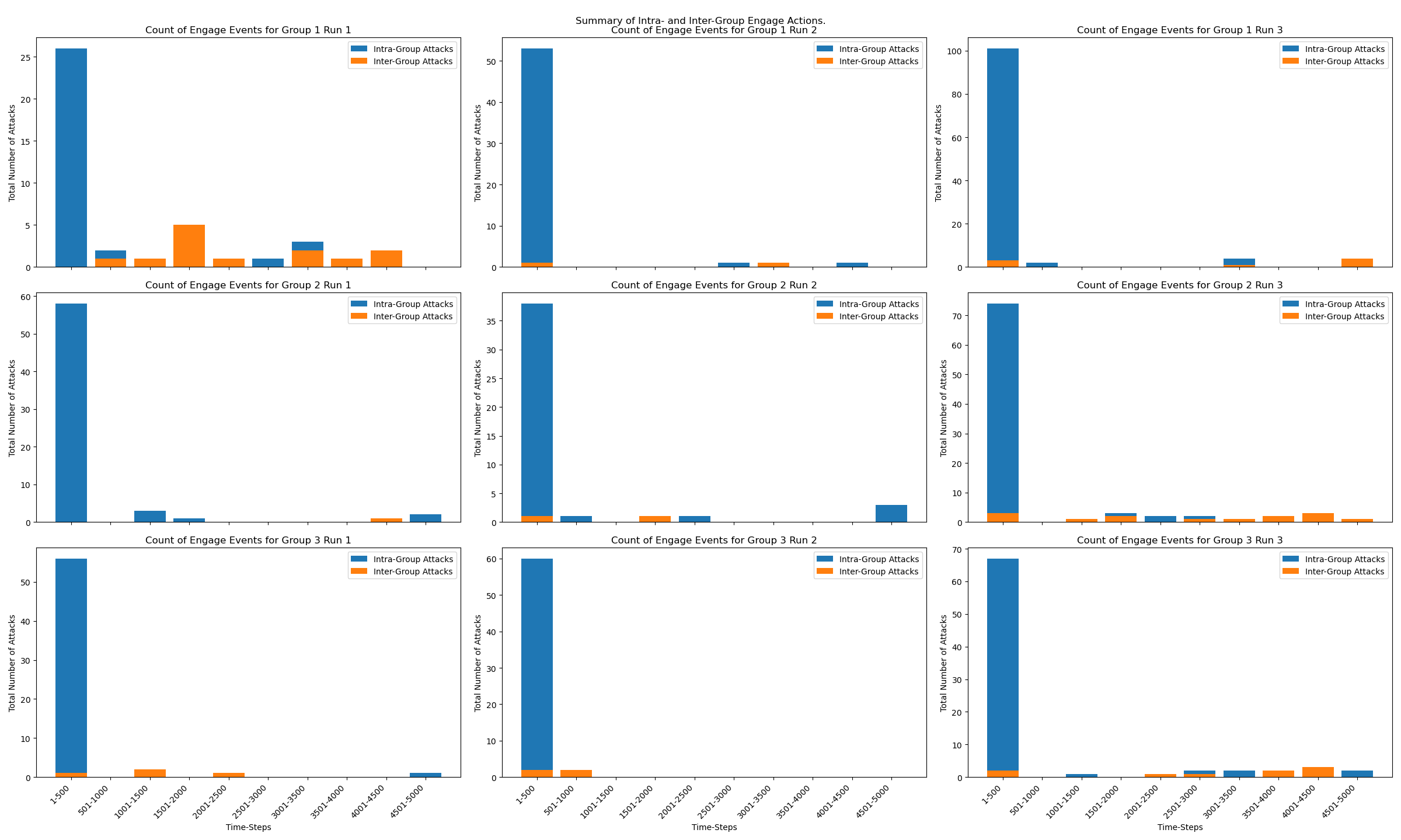}
        \caption{Summary over all engage actions that were taken in scenario 3. The orange part of the bar shows the number of engage actions that were executed between two agents of different groups. The blue part of the bar shows the number of engage actions executed group internally.}
        \label{fig:s3_engage_actionplan_summary}
    \end{figure}
    
    \begin{figure}
        \centering
        \includegraphics[height=0.85\textwidth]{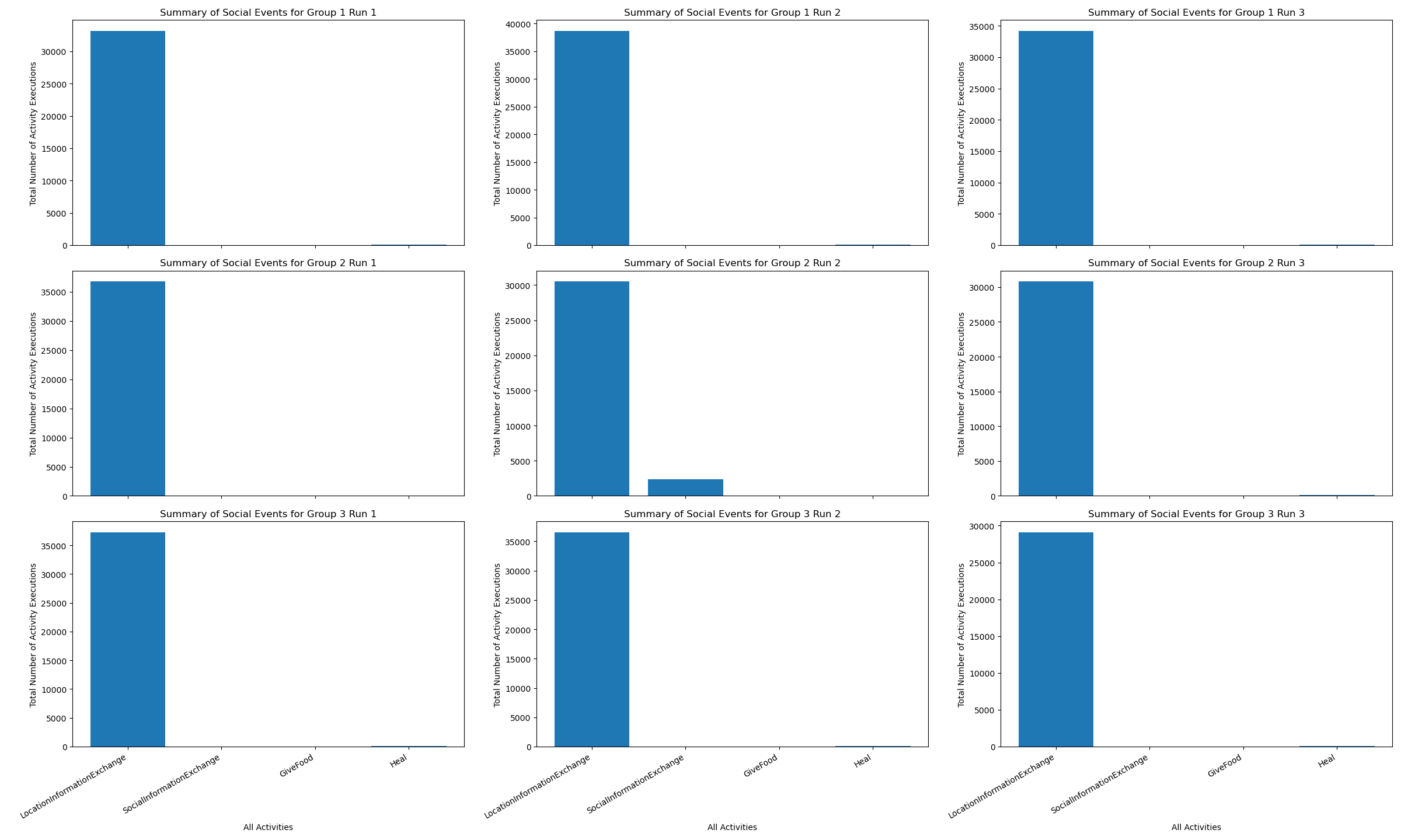}
        \caption{Count of all social actions plans for scenario 1.}
        \label{fig:s1_social_action_summary}
    \end{figure}
    
    \begin{figure}
        \centering
        \includegraphics[height=0.85\textwidth]{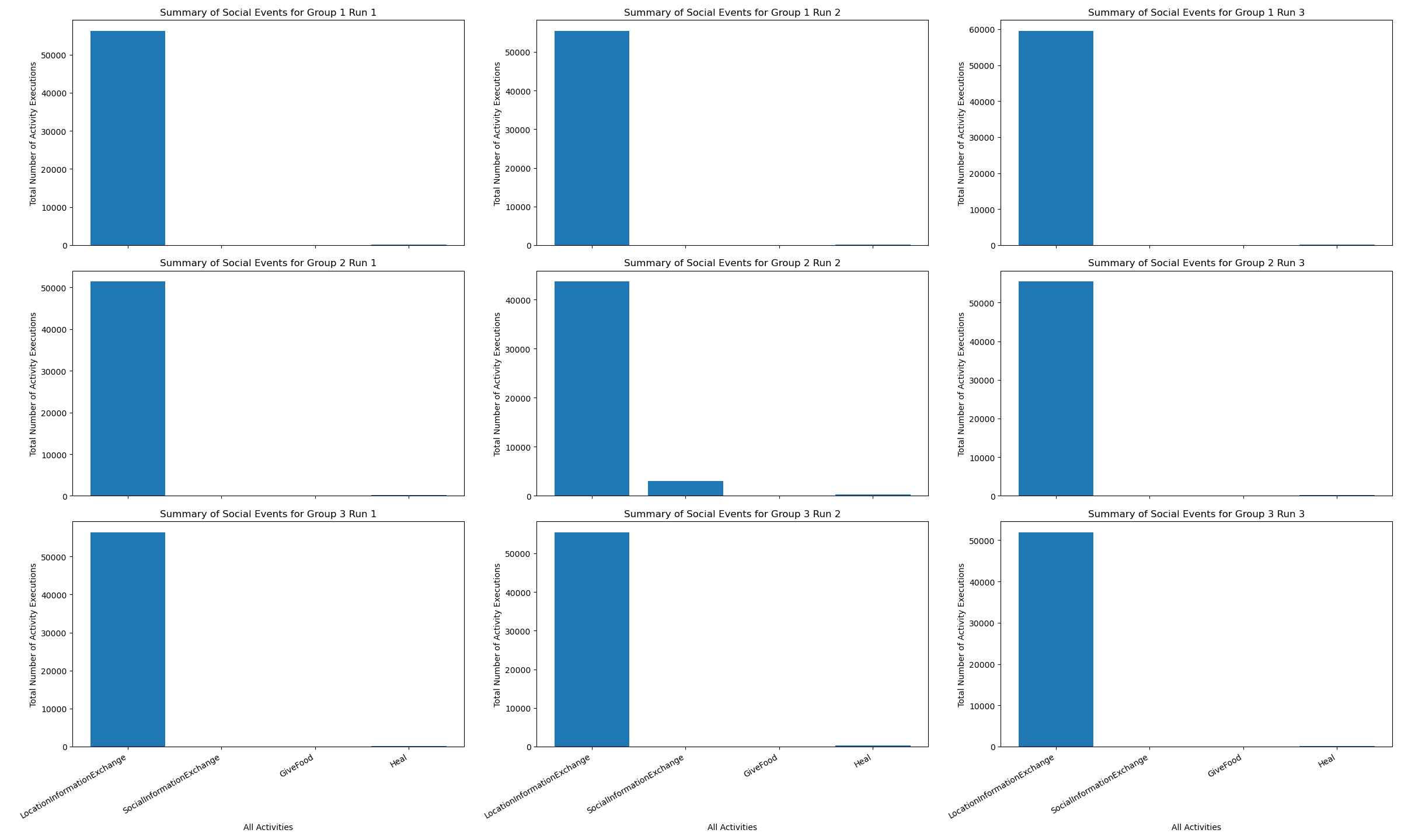}
        \caption{Count of all social actions plans for scenario 2.}
        \label{fig:s2_social_action_summary}
    \end{figure}
    
    \begin{figure}
        \centering
        \includegraphics[height=0.85\textwidth]{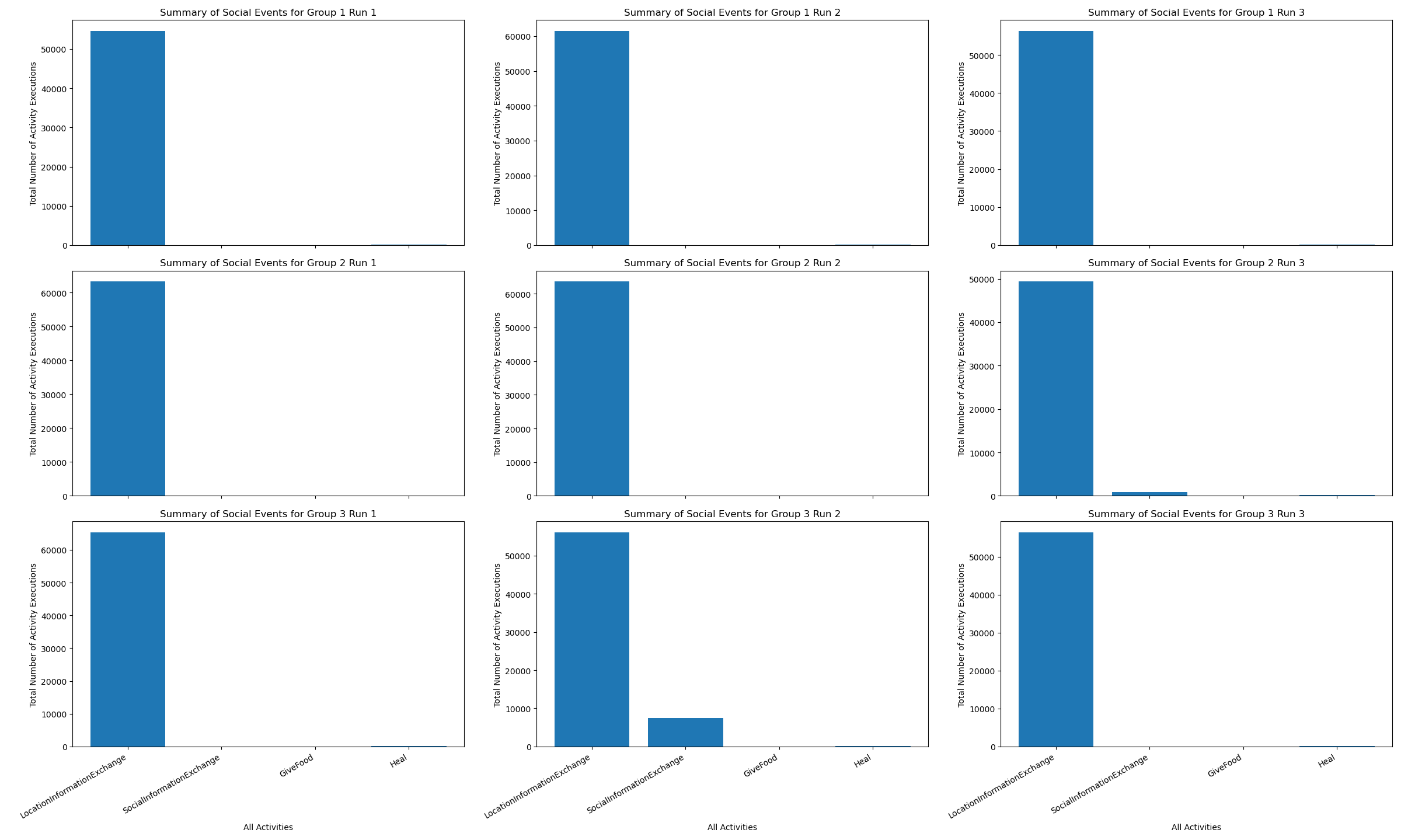}
        \caption{Count of all social actions plans for scenario 3.}
        \label{fig:s3_social_action_summary}
    \end{figure}
    
    \begin{figure}
        \centering
        \includegraphics[height=0.85\textwidth]{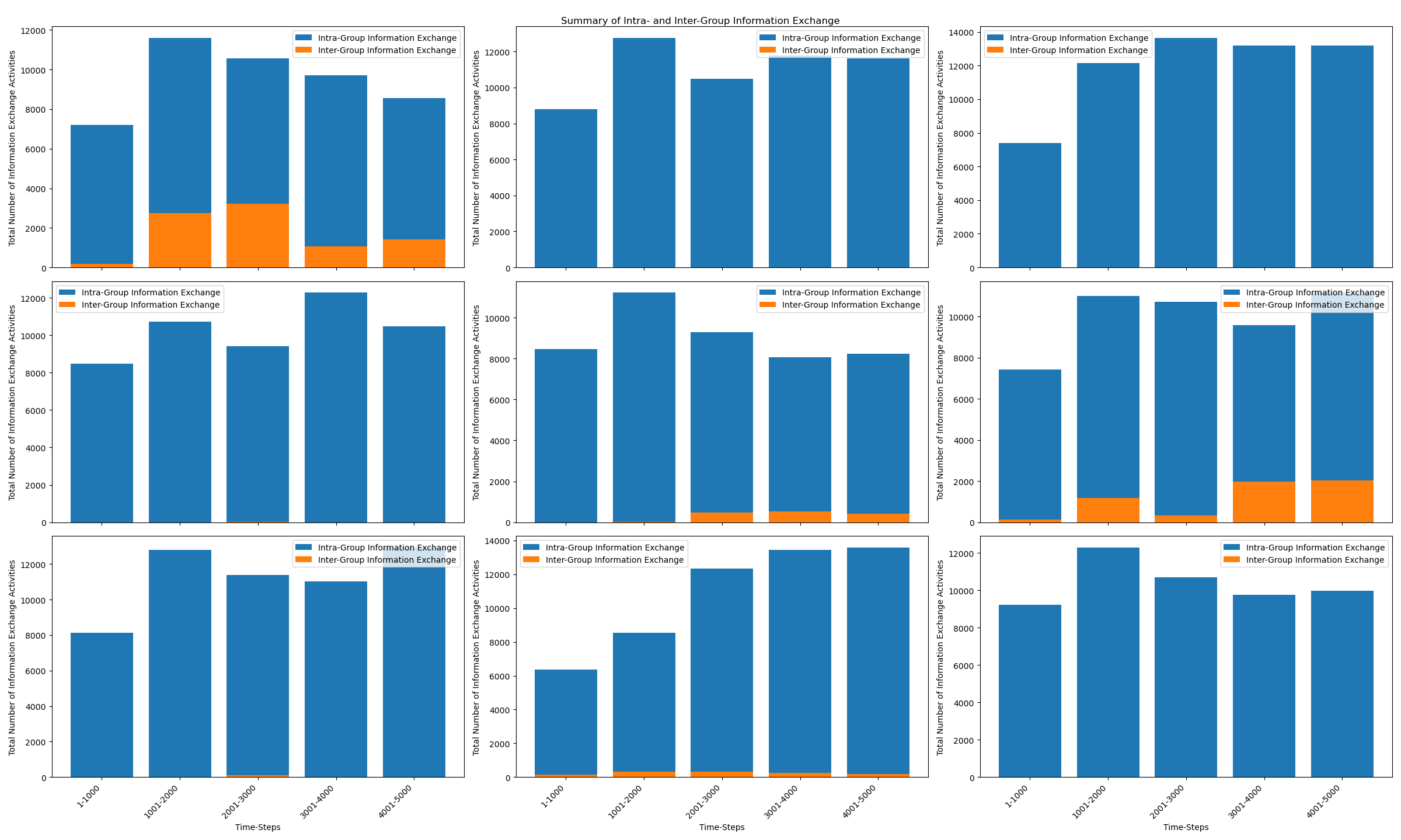}
        \caption{Count of all information exchange action plans for scenario 2. The blue part of the bar indicates intra-group social behavior while the orange part describes inter-group social behavior.}
        \label{fig:s2_information_exchange_summary}
    \end{figure}
    
    \begin{figure}
        \centering
        \includegraphics[height=0.85\textwidth]{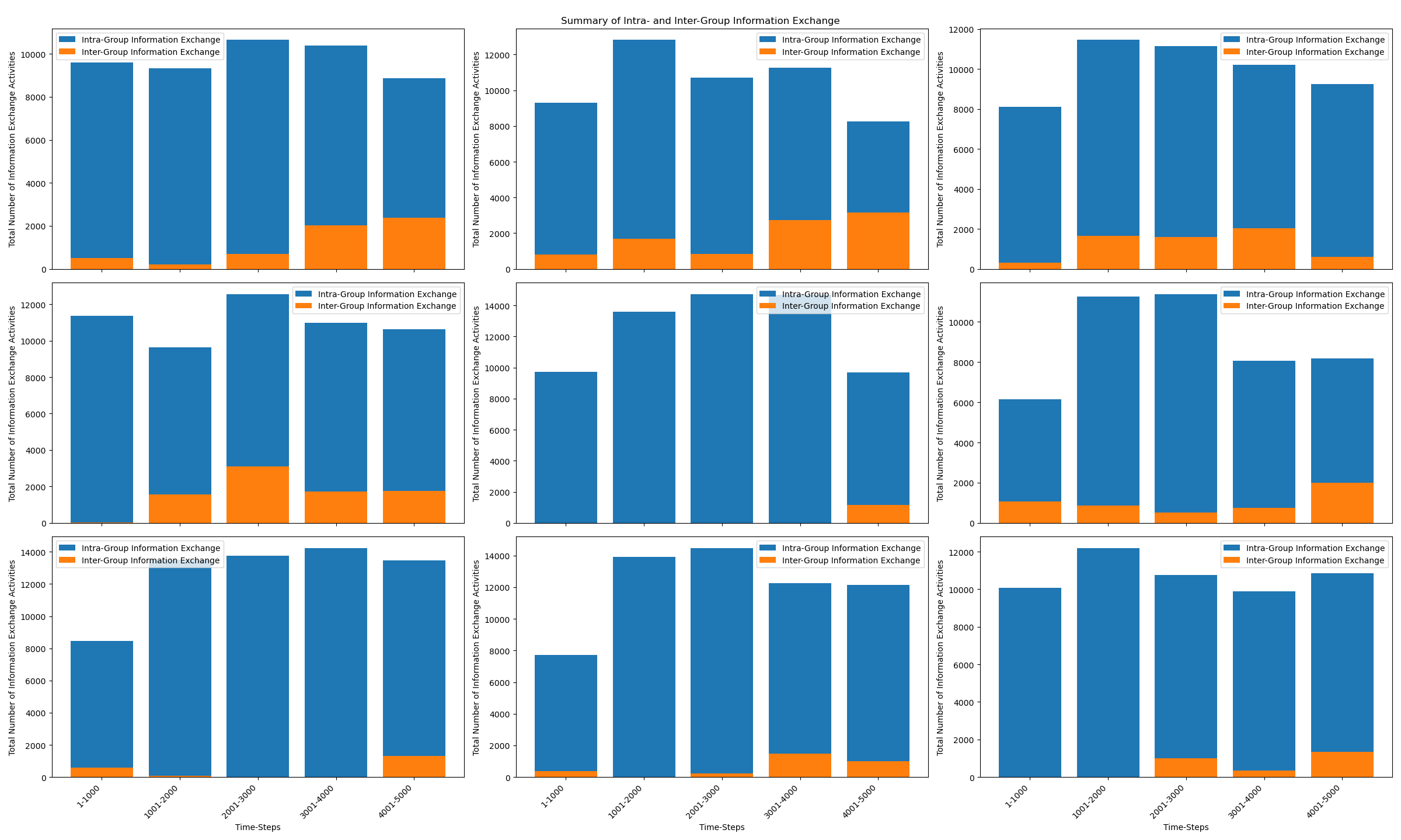}
        \caption{Count of all information exchange action plans for scenario 3. The blue part of the bar indicates intra-group social behavior while the orange part describes inter-group social behavior.}
        \label{fig:s3_information_exchange_summary}
    \end{figure}
\end{landscape}

\end{document}